\documentstyle[twocolumn,aps,prl,floats,epsf]{revtex}

\begin{document}
\epsfclipon

\draft

\wideabs{

\title{Improved Measurement of the $\bar d / \bar u$ Asymmetry in the Nucleon Sea}

\author{
R.~S.~Towell$^{a,f}$,
P.~L.~McGaughey$^f$, 
T.~C.~Awes$^i$,
M.~E.~Beddo$^h$, 
M.~L.~Brooks$^f$,
C.~N.~Brown$^c$, 
J.~D.~Bush$^a$,
T.~A.~Carey$^f$, 
T.~H.~Chang$^h$\cite{byline1},
W.~E.~Cooper$^c$,
C.~A.~Gagliardi$^j$,
G.~T.~Garvey$^f$, 
D.~F.~Geesaman$^b$, 
E.~A.~Hawker$^{j,f}$, 
X.~C.~He$^d$,
L.~D.~Isenhower$^a$,
D.~M.~Kaplan$^e$, 
S.~B.~Kaufman$^b$, 
P.~N.~Kirk$^g$, 
D.~D.~Koetke$^k$, 
G.~Kyle$^h$,
D.~M.~Lee$^f$,
W.~M.~Lee$^d$\cite{byline2}, 
M.~J.~Leitch$^f$, 
N.~Makins$^b$\cite{byline1}, 
J.~M.~Moss$^f$,
B.~A.~Mueller$^b$,
P.~M.~Nord$^k$,
V.~Papavassiliou$^h$, 
B.~K.~Park$^f$, 
J.~C.~Peng$^f$, 
G.~Petitt$^d$, 
P.~E.~Reimer$^{f,b}$,
M.~E.~Sadler$^a$,
W.~E.~Sondheim$^f$, 
P.~W.~Stankus$^i$, 
T.~N.~Thompson$^f$, 
R.~E.~Tribble$^j$,
M.~A.~Vasiliev$^j$\cite{byline3}, 
Y.~C.~Wang$^g$, 
Z.~F.~Wang$^g$, 
J.~C.~Webb$^h$, 
J.~L.~Willis$^a$,
D.~K.~Wise$^a$,
G.~R.~Young$^i$\\ \vspace*{9pt}
(FNAL E866/NuSea Collaboration)\\ \vspace*{9pt}
}
\address{
$^a$Abilene Christian University, Abilene, TX 79699\\
$^b$Argonne National Laboratory, Argonne, IL 60439\\
$^c$Fermi National Accelerator Laboratory, Batavia, IL 60510\\
$^d$Georgia State University, Atlanta, GA 30303\\
$^e$Illinois Institute of Technology, Chicago, IL  60616\\
$^f$Los Alamos National Laboratory, Los Alamos, NM 87545\\
$^g$Louisiana State University, Baton Rouge, LA 70803\\
$^h$New Mexico State University, Las Cruces, NM 88003\\
$^i$Oak Ridge National Laboratory, Oak Ridge, TN 37831\\
$^j$Texas A \& M University, College Station, TX 77843\\
$^k$Valparaiso University, Valparaiso, IN 46383\\
}
\date{\today}

\maketitle
\begin{abstract}
Measurements of the ratio of Drell-Yan yields from an
800 \rm{GeV/c} proton beam incident on liquid hydrogen and deuterium
targets are reported.  Approximately 360,000 Drell-Yan muon pairs 
remained after all cuts on the data.
>From these data, the ratio of anti-down ($\bar{d}$) to anti-up 
($\bar{u}$) quark distributions in the proton sea is determined 
over a wide range in Bjorken-$x$.  These results confirm previous measurements
by E866 and extend them to lower $x$. From these data, $(\bar{d}-\bar{u})$ 
and $\int(\bar{d}-\bar{u})dx$ are evaluated for $0.015<x<0.35$.  
These results are compared with parameterizations of various
parton distribution functions, models and experimental results from NA51, NMC and HERMES.  
\end{abstract} 
}

\section{Introduction}

Recent measurements~\cite{prl,nmc,na51,hermes} have shown
a large asymmetry in the distributions of up and down antiquarks ($\bar u$ and $\bar d$) in the
nucleon.
While no known symmetry requires $\bar u$ to equal $\bar d$, a
large $\bar d / \bar u$ asymmetry was not anticipated.  The usual
assumption was that the sea of quark-antiquark pairs is produced perturbatively from 
gluon splitting. Since the mass difference of the up and
down quarks is small, nearly equal
numbers of up and down pairs should result. Thus a large
$\bar d / \bar u$ asymmetry requires a non-perturbative origin for 
this effect.

The data from experiment E866/NuSea~\cite{prl} at Fermilab were the first 
to demonstrate a strong Bjorken-$x$ dependence of the $\bar d / \bar u$ 
ratio. In that earlier work, only data at fairly large 
dimuon mass were analyzed. In this paper we report results based on
the entire data set and describe the details of the
experimental apparatus and analysis procedure.  These data
cover a larger range of mass and Bjorken-$x$, and demonstrate 
consistency of the results for three different spectrometer settings.
They also provide more accurate determinations of $\bar d / \bar u$, 
$\bar d - \bar u$ and the integral of $\bar d - \bar u$. 
The data are compared with several parton distribution function sets,
and the implications of these results
for various models that predict a $\bar d / \bar u$ asymmetry are 
discussed.

There have been four other experimental studies~\cite{nmc,na51,hermes,e772} of the  $\bar d / \bar u$ 
asymmetry in the nucleon.  The first measurement was performed by the New Muon Collaboration (NMC).  NMC measured
the cross section ratio for deep inelastic scattering (DIS) of muons from hydrogen 
and deuterium~\cite{nmc}.  
Their extrapolated result for the integral of the difference of the proton
and neutron structure functions is
\begin{eqnarray}
\int^1_0 \left[F^{p}_2 - F^{n}_2\right] \frac{dx}{x} = 0.235 \pm 0.026.
\label{eqn:gsr}
\end{eqnarray}
This result can be compared with the Gottfried Sum Rule (GSR)~\cite{GSR}. 
The Gottfried Sum, $S_G$, can be expressed in terms of the parton distribution functions as
\begin{eqnarray}
S_G  & \equiv & \int^1_0 \left[F^{p}_2 - F^{n}_2\right] \frac{dx}{x} \\ \nonumber
    &  = & \frac{1}{3} + \frac{2}{3}\int^1_0 \left[ \bar u(x) - \bar d(x) \right] dx.
\end{eqnarray}
In the derivation of Eq. 2, charge symmetry was assumed.
If it is also assumed that $\int \bar{d} (x) dx = \int \bar{u} (x) dx$, then one arrives 
at a GSR result of $1/3$, in disagreement with the NMC result.  Rather,
the NMC measurement implies
\begin{eqnarray}                                                 
\int^1_0 [ \bar d(x) - \bar u(x)]dx = 0.148 \pm 0.039.
\end{eqnarray}
The NMC measurement~\cite{nmc} was the
first indication that there are more anti-down quarks in the proton than
anti-up quarks.
  
In order to obtain the Gottfried Sum from the NMC data, an extrapolation was needed
to account for contributions to the sum for $x \le 0.004$.
Since $F_2/x$ rises rapidly in this region, a sizable
contribution to $S_G$ was expected.
The small-$x$ extrapolation was checked by Fermilab E665~\cite{E665}, which made a similar 
measurement as NMC except that they measured the ratio for \mbox{$10^{-6}\le x \le 0.3$}. 
Over the kinematic range where NMC and E665 overlap, their measurements 
agree.  However, E665 determined that for $x \le 0.01$ the value of  
$2F^d_2/F^p_2 -1$ was a constant $0.935 \pm 0.008 \pm 0.034$.  
While this could be interpreted as a difference between $F^n_2$ and $F^p_2$,
it is usually thought to be the effect of nuclear shadowing in deuterium~\cite{shadow,Bade94}
which means that $F^n_2/F^p_2 \ne 2F^d_2/F^p_2 -1$.
Therefore it is difficult to measure $F^n_2/F^p_2$ in a 
model-independent way at low $x$.

Following the publication of the NMC result, it was suggested~\cite{ES} 
that the Drell-Yan process~\cite{DY} could provide 
a more direct probe of the light antiquark asymmetry of
the nucleon.  
In the parton model, the Drell-Yan cross section at leading order is 
\begin{eqnarray}
\lefteqn{\frac{d^2\sigma}{dx_1 dx_2} =} \hspace{0.5in}\\ \nonumber
                         & & \frac{4 \pi \alpha ^2}{9 M^2}
                             \sum_i e_i^2 \left[f_i(x_1)\bar f_i(x_2) +
                             \bar f_i(x_1) f_i(x_2)\right],
\label{eqn:dy}
\end{eqnarray}
where the sum is over all quark flavors, $e_i$ are the quark charges, $f_i$ are
the parton distribution functions,  and $M$ is the virtual photon or
dilepton mass~\cite{DYeqn}.  Here  $x_1$ and $x_2$ are the Bjorken-$x$ 
of the partons from the beam and target, respectively.

Two kinematic quantities commonly used to describe Drell-Yan events are the 
Feynman-$x$ ($x_F$) and the dilepton mass ($M$) which are defined as :

\begin{equation}
x_F = \frac{p^{\gamma}_{||}}{p^{\gamma,max}} \approx \frac{p^{\gamma}_{||}}{\sqrt{s}/2} = x_1 - x_2
\end{equation}
and
\begin{equation}
M^2 = Q^2 \approx x_1 x_2 s,
\end{equation}
where $p^{\gamma}_{||}$ is the center-of-mass longitudinal momentum of the 
virtual photon, $p^{\gamma,max}$ is its maximum possible value, and $s$ is 
the total four-momentum squared of the initial nucleons.  
The proton-deuterium Drell-Yan cross section can be expressed as
\begin{eqnarray}
\sigma^{pd} \approx \sigma^{pp} + \sigma^{pn},
\label{eqn:depn}
\end{eqnarray}
which ignores the small nuclear effects inside the deuterium nucleus.  
Using this approximation and assuming charge symmetry, 
the cross section ratio for Drell-Yan events produced in
deuterium and hydrogen targets can be used to determine the ratio
$\bar d / \bar u$. 

The first experiment to use this idea was
the NA51 experiment~\cite{na51} at CERN.  This experiment used the 
$450$~\rm{GeV/c} proton beam from the CERN-SPS with liquid hydrogen
and deuterium targets.  
The NA51 experiment was able to reconstruct almost 6,000 Drell-Yan events with 
the dimuon mass above $4.3$~\rm{GeV/c$^2$}, and from these data
they obtained
\begin{eqnarray}
\left. \frac{\bar{d}}{\bar{u}} \right|_{\langle x \rangle=0.18}
 = 1.96\pm 0.15\pm 0.19.                                   
\end{eqnarray}
However, the NA51 spectrometer's acceptance
was peaked near $x_F = 0$ and  $x = 0.18$.  This, combined
with their limited statistics, made it impossible to determine 
the $x$-dependence of the ratio.

Several groups have performed global fits to existing data
from DIS, Drell-Yan, and other processes
to generate parameterizations of parton distribution
functions (PDFs)~\cite{CTEQ,MRS,grv95,pdflib}. 
Prior to the measurements by NMC and NA51 
the usual assumption was that $\bar {d}(x) = \bar {u}(x)$.  
The PDFs were then revised to accommodate the NMC and NA51 data.
While these measurements show that $\bar d \ne \bar u$, neither imposed rigid
constraints on the $x$-dependence of the $\bar d(x)$/$\bar u(x)$ 
asymmetry. 

A better
measurement of $\bar d/\bar u$ is possible with Drell-Yan if the detector acceptance is 
largest for $x_F > 0$, since the Drell-Yan cross section ratio is more
sensitive to the target antiquark distribution in this kinematic regime.
This increase in  sensitivity results from the Drell-Yan cross section 
being dominated by the annihilation 
of a beam quark with a target antiquark in this kinematic regime.
For $x_1 \gg x_2$,  one obtains
\begin{eqnarray}
\sigma^{pp} \propto \frac{4}{9} u(x_1)\bar u(x_2) + \frac{1}{9} d(x_1)\bar d(x_2)
\label{eqn:sigpp2}
\end{eqnarray}
and
\begin{eqnarray}
\sigma^{pn} \propto \frac{4}{9} u(x_1)\bar d(x_2) + \frac{1}{9} d(x_1)\bar u(x_2).
\label{eqn:sigpn2}
\end{eqnarray} 
>From Eqs. \ref{eqn:depn}, \ref{eqn:sigpp2}, and \ref{eqn:sigpn2} it is a simple matter to derive 
\begin{equation}
\left. \frac{\sigma^{pd}}                            
           {2\sigma^{pp}}
\right|_{x_1\gg x_2} \approx\frac{1}{2}
\frac{\left[ 1 + \frac{1}{4}\frac{d(x_1)}{u(x_1)} \right]}
     {\left[ 1 + \frac{1}{4}\frac{d(x_1)}{u(x_1)}
           \frac{\bar{d}(x_2)}{\bar{u}(x_2)} \right]}
      \left[ 1 + \frac{\bar{d}(x_2)}{\bar{u}(x_2)} \right].
\label{eqn:xf1}
\end{equation}
This expression can be further simplified since $d(x) \ll 4u(x)$, 
resulting in
\begin{equation}                                                            
\left. \frac{\sigma^{pd}}
           {2\sigma^{pp}}
\right|_{x_1\gg x_2} \approx\frac{1}{2}
      \left[ 1 + \frac{\bar{d}(x_2)}{\bar{u}(x_2)} \right].
\label{eqn:prl41}                                           
\end{equation}  
This equation illustrates the sensitivity of the Drell-Yan cross 
section ratio to $\bar d/\bar u$ for $x_1 \gg x_2$.

In FNAL E866/NuSea \cite{prl} the ratio of the Drell-Yan 
cross section for proton-deuteron interactions to that for proton-proton interactions
was measured over a wide range of $x$ and other kinematic variables.  
This measurement in turn provided an accurate determination of 
 $\bar d(x)$/$\bar u(x)$ and an independent
determination of the integral of \hbox{$\left[\bar d(x)-\bar u(x)\right]$} over the same
$x$ region.

Recently, the HERMES collaboration \cite{hermes} has reported a measurement of $\bar{d}-\bar{u}$
over the range $0.02 < x < 0.30$, based on a measurement of semi-inclusive deep-inelastic scattering.
The HERMES results are in good agreement with the results from FNAL E866/NuSea, but have limited 
precision.

In Ref. \cite{prl}, we presented initial results of the FNAL E866/NuSea study of the light 
antiquark asymmetry in the nucleon sea, based on an analysis of approximately 40\% of our data.
Here we present the final results of the analysis of the full data set from the experiment.

\section{Experimental Setup}

FNAL E866/NuSea used an 800~\rm{GeV/c} proton beam extracted
from the Fermilab Tevatron accelerator and transported to 
the east beamline of the Meson experimental hall.  
The beam position and shape were measured using RF cavities and segmented-wire 
ionization chambers (SWICs). The final SWIC was located 1.7~m  upstream
of the target.
The beam at this SWIC was typically 6~\rm{mm} wide and 
1~\rm{mm} high (FWHM).  
The most important beam intensity measurement 
was made with a secondary emission monitor (SEM) located about 100~m upstream of the targets. 
In addition to the SEM, the beam intensity was monitored with a quarter-wave 
RF cavity and an ionization chamber.
The nominal beam intensity ranged
from $5 \times 10^{11}$ to $2 \times 10^{12}$
protons per 20 second spill, depending on the spectrometer magnet setting.

The proton beam passed through one of three physically identical, 
thin, stainless steel target flasks.  
These flasks were cylindrical in shape with hemispherical ends 
and insulated vacuum jackets.
The flasks were 7.62 \rm{cm} in diameter and 50.8 \rm{cm} in length.  
The two end windows on each flask totaled
0.10 \rm{mm} of stainless steel and 0.28 \rm{mm} of titanium. One flask was filled
with liquid deuterium, another was filled with liquid hydrogen, and the third was 
evacuated.  
The hydrogen target was
7\% of an interaction length and 6\% of a radiation length,
and the deuterium target was
15\% of an interaction length and 7\% of a radiation length.
The evacuated target was less than 0.2\%
of an interaction length and 1.4\% of a radiation length.  
Both the temperatures and vapor pressures of the 
filled flasks were monitored.  

All three flasks were mounted on a movable table so that the target could be
changed during the 40 second gap between the 20 second 
beam spills.  The normal target cycle consisted of twelve 
spills with five spills on the deuterium target, one spill 
on the empty flask, five spills on the hydrogen 
target and another spill on the empty flask.  
This frequent cycling of the targets minimized many 
systematic uncertainties.

At $85^\circ$ to the beam direction there
were a pair of four-element scintillator telescopes.
These viewed the target through a hole
in the heavy shielding enclosing the target area to monitor the luminosity, 
duty factor, data-acquisition live time,
and to independently verify which target was in the beam.
 
\begin{figure*}                                                     
  \begin{center} 
    \mbox{\epsfxsize=5.25in\epsffile{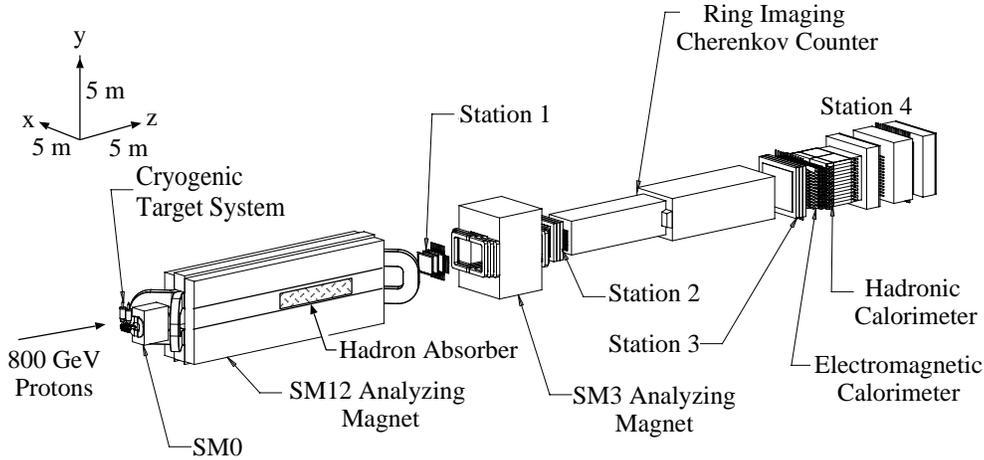}}
  \end{center}       
  \caption{The FNAL E866/NuSea Spectrometer}             
  \label{fig:spec}
\end{figure*}

The detector apparatus used in this experiment was the E605 dimuon 
spectrometer~\cite{moreno}, shown in Fig. \ref{fig:spec}.
While changes were made to the 
spectrometer for E866, the basic design has remained the same since the 
spectrometer was first used for E605 in the early 1980's.  
Three large dipole magnets provide for the momentum analysis of energetic
muons, while deflecting soft particles out of the acceptance. The magnetic
fields are in the horizontal direction, bending the tracks in the vertical direction.
The polarities and currents of the first two magnets were adjusted to
select particular ranges of dimuon mass, while minimizing background rates
in the drift chambers.
The changes to the spectrometer for E866 were the installation
of six new drift chamber planes at the first tracking station, a reconfigured
absorber wall, two new hodoscope planes~\cite{towell}, and a new trigger 
system~\cite{trig}.

The first dipole magnet (SM0) was used
to increase the opening angle of muon pairs when taking data with 
the magnets configured to have acceptance at the lowest mass.
For the higher mass settings it was not energized.

A water-cooled copper beam dump was located at a distance of 1.73 m into the second magnet~(SM12).
The protons that passed through the target
were absorbed in the 3.28-m-long dump.  The beam dump was
about 22 interaction lengths, or 230 radiation lengths, thick.
It filled the magnet aperture in the horizontal
direction for most of its length, but was a maximum of 25.4~\rm{cm} 
high in the vertical direction.  This allowed many of the 
muons of interest to travel above or below the beam dump, 
minimizing muon multiple scattering and energy loss.

Downstream of the beam dump was an absorber wall 
that completely filled the aperture
of the magnet.  This wall consisted of
0.61 m of copper followed by 2.74 m of carbon and
1.83 m of borated polyethylene.  
The effect of this wall, which was over thirteen interaction lengths and
sixty radiation lengths long, was to absorb 
most of the produced hadrons, electrons, and gammas.
Effectively only muons traversed the active elements of the 
spectrometer, allowing 
the use of high beam intensities while keeping the instantaneous number of hits in 
each drift chamber at an acceptable level.

The third magnet~(SM3), located downstream of SM12 and the first tracking
station, provided the magnetic field used for the momentum determination of the muons. 
The position of each muon was measured precisely at three tracking
stations, one upstream and two downstream of SM3.
Each tracking station consisted of three 
pairs of high-rate drift chambers, followed by
horizontal and vertical scintillation hodoscopes 
used to generate the dimuon trigger.  
(The exception to this configuration was the absence of the  
hodoscope that provides horizontal position information 
after the second tracking station. 
This hodoscope was omitted to minimize multiple scattering
between the second and third tracking stations.)

At the end of the spectrometer, behind shielding,
was the fourth tracking station. It consisted of three planes of proportional tubes and a pair
of hodoscope planes. The ring imaging Cherenkov counter (RICH) and two calorimeters, shown in 
Fig.~\ref{fig:spec}, were not active in E866.   The RICH was filled with 
helium to reduce multiple scattering between the second and third tracking 
stations.
Summaries of the physical construction of the drift chambers, hodoscopes, and 
proportional tubes may be found in Ref. \cite{towell}.

\section{Trigger and Monitoring}

The trigger was optimized to detect dimuon events originating from the target, 
while rejecting as many muons produced in the beam dump as possible. 
A new trigger system was implemented for E866~\cite{trig,hawker}. 
It used the hodoscope signals to determine whether the event should be written 
to tape.  
The hits in the hodoscopes at stations 1, 2, and 4 that measured the vertical 
track positions were compared with the contents of 
a three-dimensional look-up table.  This table was generated by Monte Carlo studies 
of dimuon events from the target.  When the hits in the scintillators matched one of
the pre-calculated dimuon trajectories, the trigger fired.  

In addition to the standard physics triggers optimized to detect 
oppositely charged dimuon events from the targets, other triggers were prescaled
to record a limited number of study events.  These study events included 
single-muon events, events satisfying triggers that relied only on the hodoscope 
planes that provided horizontal position information, 
and other diagnostic triggers such as 
two like-sign muons from the target area 
that traveled down opposite sides (left and right) of the spectrometer.  

For each 20 second beam spill, information important for analysis 
was recorded as part of the data stream.  Beam intensity, position, size, and duty
factor were recorded, as well as the pressure, 
temperature, and positions of the liquid targets, magnet voltages and 
currents, and various monitors used for calculating the readout deadtime. 
The beam position and size were stable throughout the experiment, well
within the dimensions of the target flasks.

\begin{table}[tbp]
\caption{Average trigger rates per beam spill and live times for the deuterium target.}
\label{tab:trigger}   
\begin{center}                                                          
\begin{tabular}{ccc}               
 mass setting	 & triggers/spill & live time     \\ \hline 
 low	      & 2200 &   99.0\%   \\ 
 intermediate         & 3200 &   97.9\%   \\
 high  	      & 2100 &   98.5\%   \\
\end{tabular}
\end{center}
\end{table}

To better monitor the spectrometer performance and data quality, 
a portion of the data was analyzed in real time. The efficiency of
each detector element and the overall track reconstruction efficiency
were carefully studied. The wire chambers had average efficiencies of 96\%.
The individual hodoscopes used in the trigger were 99\% efficient.
The overall trigger efficiency was greater than 94\%.  
Average trigger rates and live times for the deuterium
target for the three spectrometer settings are given in Table \ref{tab:trigger}.  
Trigger rates were lower and live times higher for the hydrogen target
(not shown). 

\section{Analysis}

\begin{figure}
  \begin{center}
    \mbox{\epsffile{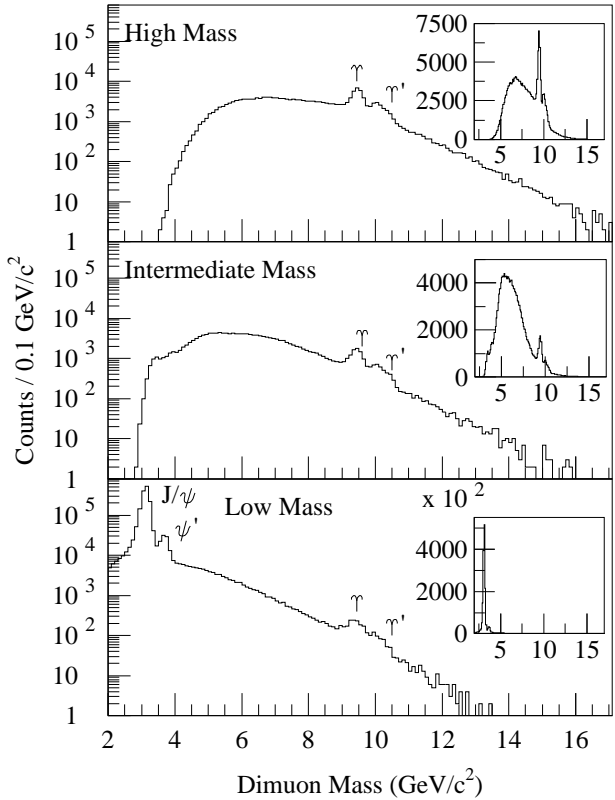}}
    \vspace*{-0.05in}                                
  \end{center}       
  \caption{The dimuon mass distributions for the three different mass settings.
		The inset figures are the same spectra shown on linear scales.
		The mass cuts used in the analysis to select Drell-Yan events 
		are listed in Table \ref{tab:massregions}.}
  \label{fig:3mass}                    
\end{figure}      

The data were taken with three mass settings of the spectrometer magnets, 
designated as the high, intermediate, and low mass settings.
Figure \ref{fig:3mass} shows the 
dimuon mass distributions for the three mass settings.
The data were further divided based upon the magnet 
polarity and deuterium target purity.  
Six data sets contained data 
useful for this analysis and are summarized in Table \ref{tab:datasets}.

\begin{table}
\caption{Summary of the data sets.  The size of each set is shown
as the number of fully reconstructed Drell-Yan events rounded to the nearest thousand.  
Magnet p$_{T}$ kicks are given for SM0 and SM12.  SM3 always provided an average
p$_{T}$ kick of 0.9~GeV/c with the same polarity as SM12.  All fields are known to $\pm 2\%$.
The uncertainties on the deuterium purity are given in Table~\ref{tab:ld22}.} 
\label{tab:datasets}   
\begin{center}                                                          
\begin{tabular}{cccc}               
 mass	 & Drell-Yan & SM0/SM12    & deuterium \\
 setting & events    & $\langle p_{T} \rm{kick} \rangle $ [GeV/c]           &  purity \\ \hline
 low	      & 89 k &    -1.04 / ~4.72    & 99.99\% \\ \hline
 intermediate & 78 k &     ~~~0~ / ~4.72    & 99.99\% \\
 	          & 50 k & ~~~0~ / -4.72    & 99.99\% \\ \hline
  	          & 37 k & ~~~0~ / ~6.39    & 99.99\% \\
 high  	      & 80 k &     ~~~0~ / ~6.39    & 97.0~\% \\
              & 24 k &     ~~~0~ / -6.39    & 97.0~\% \\
\end{tabular}
\end{center}
\end{table}

\begin{table}
\caption{Mass regions used for each spectrometer setting for 
Drell-Yan analysis.}
\label{tab:massregions}   
\begin{center}                                                          
\begin{tabular}{ccc}               
 mass setting	 & mass regions accepted    \\ \hline
 low	      & 4.0 to 8.8 \rm{GeV/c$^2$}   \\ 
 intermediate         & 4.3 to 8.8 \rm{GeV/c$^2$} and $> 10.8$ \rm{GeV/c$^2$}   \\
 high  	      & 4.5 to 9.0 \rm{GeV/c$^2$} and $> 10.7$ \rm{GeV/c$^2$} \\
\end{tabular}
\end{center}
\end{table}

A first-pass analysis of the data was done on Fermilab's IBM
parallel-computing UNIX farms.  
Since only about 1\% of the events written to tape 
reconstructed to form a dimuon event from the target, this 
analysis efficiently reduced the raw data tapes
to a small number of data summary tapes (DSTs). 
After the individual tracks were fully reconstructed, muon pairs were 
identified.  Fewer than 0.08\% of all the fully reconstructed events contained more 
than two muon tracks from the target, resulting in virtually 
no combinatorial ambiguities. 

A second-pass analysis of the DSTs was performed with many small changes to 
optimize the mass resolution and to study systematic effects.
The results were written to PAW ntuples~\cite{paw} for physics analysis.  

Final cuts on the data were carefully studied to assure the removal of
bad events, such as interactions outside of the target
region.  Events were also cut if the
reconstructed tracks did not satisfy the trigger conditions.
Each beam spill was required to meet 
certain quality criteria.  The beam duty factor, readout live time, and beam 
intensity were all required to exceed minimum values.

\begin{figure}
  \begin{center}
    \mbox{\epsfxsize=8.5cm\epsffile{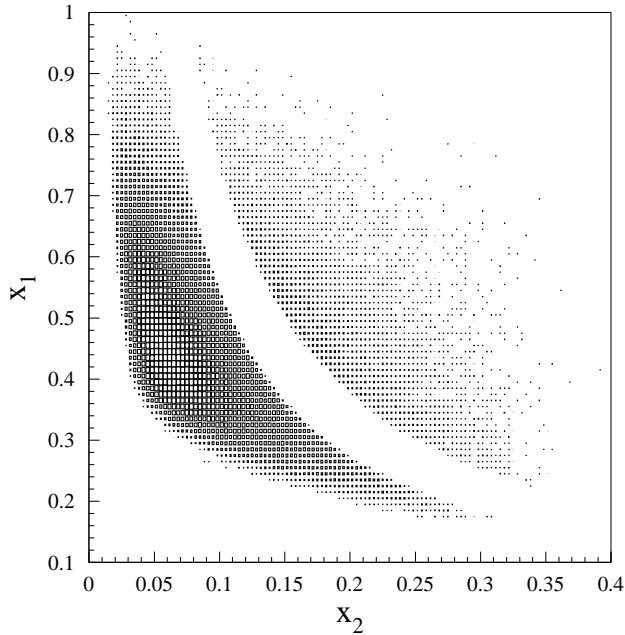}}
    \vspace*{-0.05in}                                
  \end{center}       
  \caption{The dimuon distributions for $x_1$ versus $x_2$ for the high mass setting.}
  \label{fig:hix1x2}                    
\end{figure}

\begin{figure}[b!]
  \begin{center}
    \mbox{\epsfxsize=8.5cm\epsffile{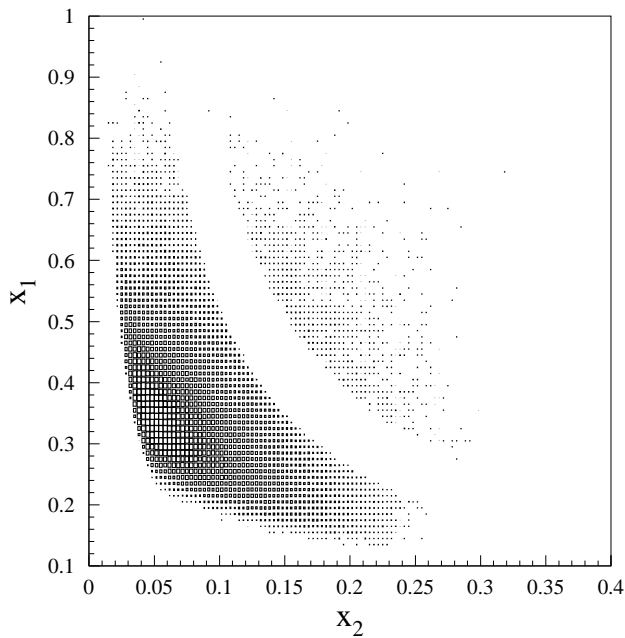}}
    \vspace*{-0.05in}                                
  \end{center}       
  \caption{The dimuon distributions for $x_1$ versus $x_2$ for the 
  intermediate mass setting.}
  \label{fig:intx1x2}                    
\end{figure}

\begin{figure}
  \begin{center}
    \mbox{\epsfxsize=8.5cm\epsffile{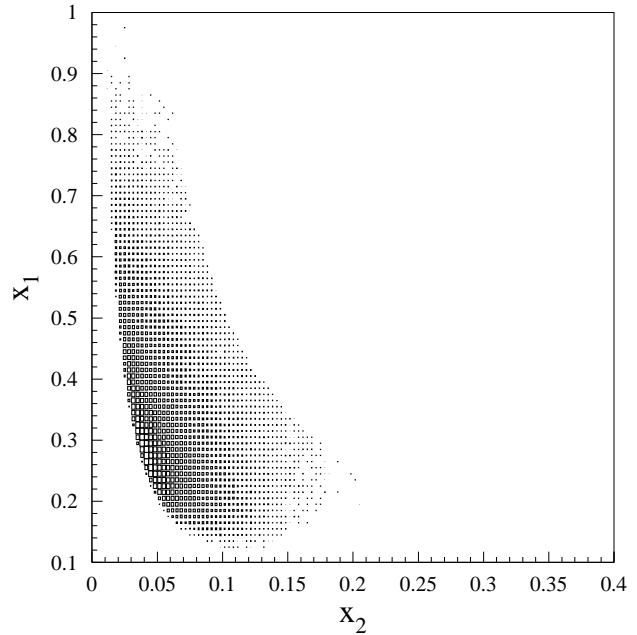}}
    \vspace*{-0.05in}                                
  \end{center}       
  \caption{The dimuon distributions for $x_1$ versus $x_2$ for the low mass setting.}
  \label{fig:lowx1x2}                    
\end{figure}

A dimuon mass cut was used to remove the $J/\psi$ and $\Upsilon$ 
resonance families from the Drell-Yan continuum.\footnote{The typical one standard deviation mass
resolution at the $J/\psi$ was 100 MeV/c$^2$ and at the $\Upsilon$ was 150 MeV/c$^2$.} 
The mass regions 
used for each data set are given in Table  \ref{tab:massregions}.
The number of events remaining in 
each of the data sets is shown in Table \ref{tab:datasets}.
Figures \ref{fig:hix1x2}, \ref{fig:intx1x2} and \ref{fig:lowx1x2}  show the 
resulting dimuon distributions for the three mass settings 
versus $x_1$ and $x_2$.

An important background was the random coincidence of two unrelated, 
oppositely charged muons.
These events are referred to as randoms.  
The data were corrected for random dimuons by subtracting 
normalized samples of pairs of combined single muon events from the dimuon sample.
The normalization was obtained from the measured yield of like-sign dimuons.
The kinematics of the like-sign events were converted to those of opposite-sign 
pairs by reflecting the vertical angle of one of the tracks, 
which is equivalent to switching the charge of that muon.
There was excellent agreement between the kinematic distributions of
these simulated random dimuons and the measured like-sign pairs after reflection.
Since most of the combined singles events reconstructed to a low
effective dimuon mass, the randoms correction was largest 
in the low-mass data.  

The average randoms correction for each mass setting is shown in
Table \ref{tab:randoms}. Estimates of single muon rates from J/$\psi$ 
and semi-leptonic 
charm decay, folded with the detector acceptance,  are consistent with the 
observed number of randoms. Another possible background is 
the dual semi-leptonic decay of $c \bar c$ to a correlated $\mu^+ \mu^-$.
However, both the mass and acceptance for these muon pairs are very low,
leading to a negligible rate in the Drell-Yan mass regions selected
above. 

\begin{figure}
  \begin{center}
    \mbox{\epsfxsize=8.5cm\epsffile{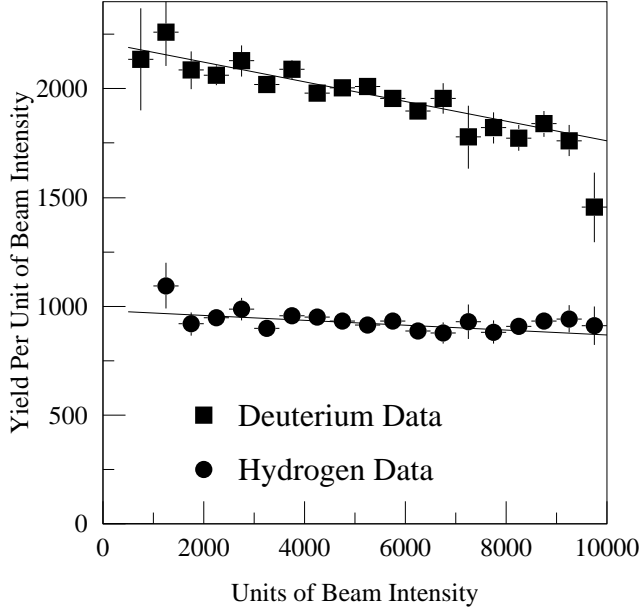}}
    \vspace*{-0.05in}                                
  \end{center}       
  \caption{The rate dependence of the low-mass data.  
           The yield of Drell-Yan events per unit of beam intensity
	   is shown versus the beam intensity for both the hydrogen and deuterium events
	   after corrections due to readout deadtime have been made.
	   The solid lines are a linear fit to the data.}
  \label{fig:rate_problem}                    
\end{figure}

A rate-dependence correction was made for the inefficiency
in event  detection and reconstruction that
occurred as a function of beam intensity.  The primary source of this
inefficiency is believed to be drift chamber hits lost
due to pileup in the single hit TDCs.
A decrease in reconstruction efficiency 
is clearly seen in 
the low-mass data shown in Fig.~\ref{fig:rate_problem}.
The yield of Drell-Yan events per unit beam intensity decreases
as the beam intensity increases.  

In order to correct the data, the reconstruction efficiency as a 
function of the beam intensity must be determined. 
Fits were made to the event yield, normalized by the beam intensity,
versus intensity.  
The data suggest that the reconstruction efficiency drops 
in a linear manner, and this basic assumption was justified by extensive 
Monte Carlo simulations.  
The reconstruction efficiency function was determined independently for each mass 
setting.
The important quantity is not the absolute 
rate dependence inefficiency, but rather the difference between the 
inefficiencies for the hydrogen and deuterium targets.  
The fits to the low mass data 
are shown in Fig. \ref{fig:rate_problem}.
The final correction to $\sigma^{pd}/2\sigma^{pp}$ due to the rate dependence
is given in Table \ref{tab:randoms}.
Another concern was that the rate dependence might also be a function
of the kinematics of the dimuon event.  This dependence was 
not observed in either the data or Monte Carlo events.

\begin{table}
\caption{Size of the randoms (background) correction for each mass setting
and correction to $\sigma^{pd}/2\sigma^{pp}$ due to the rate-dependence
effect.} 
\label{tab:randoms}
\begin{center}                                                          
\begin{tabular}{cccc}               
mass        	& \% random & $\langle$mass$\rangle $     	& rate correction \\ 
setting	   	&  events   &  (randoms)        & to $\sigma^{pd}/2\sigma^{pp}$ \\ \hline
low	        & $4.1$\% 	&  4.5 \rm{GeV/c$^2$}	    & $5.45\%	\pm 0.82\%$ \\ 
intermediate	& $2.9$\% 	&  5.1 \rm{GeV/c$^2$}     & $1.06\%	\pm 0.89\%$ \\ 
high		    & $0.2$\%  	&  5.4 \rm{GeV/c$^2$} 	& $1.76\%   \pm 0.69\%$ \\ 
\end{tabular}
\end{center}
\end{table}
 
The data included in this analysis were taken over
a period of five months.  
The deuterium target was filled twice during this time.  The analysis of 
the first fill indicated that the deuterium purity was 99.99\%.  
The second fill was of a slightly lesser quality.
Table \ref{tab:ld22} shows the composition 
of the second deuterium fill, based on two independent assays. 
The purity of the liquid hydrogen target was better than 99.99\%.

\begin{table}
\caption{Composition of the second 
	 deuterium fill.  The results shown are in percent volume.} 
\label{tab:ld22}
\begin{center}                                                          
\begin{tabular}{cc}               
material	& percent volume   \\ \hline
D$_2$		& $94.05\% \pm 0.6\%$ \\ 
HD		& ~$5.90\%	\pm 0.6\%$  \\ 
H$_2$		& ~~$0.05\% \pm 0.01\%$  \\ \hline 
deuterium	& ~$97.0\% \pm 0.6\%$   \\
hydrogen	& ~$3.0\%  \pm 0.6\%$   \\
\end{tabular}
\end{center}
\end{table}
 
The density of the target material was determined from
the vapor pressure of the gas above the liquid in both cryogenic systems.
These pressures were constantly monitored and recorded in a database.
The temperature of each flask was also recorded.  From these data
the average pressure was determined 
for each target and for each data set.  These averaged close to
15 psi. Cryogenic data tables~\cite{ld2} for hydrogen and deuterium were
used to convert the vapor pressures to the mass densities shown
in Table \ref{tab:density}.

\begin{table}
\caption{Average density in \rm{g/cm$^3$} of the liquid targets for each data set.} 
\label{tab:density}
\begin{center}                                                          
\begin{tabular}{cccc}               
 mass	       & SM0/SM12    & hydrogen          & deuterium  \\
 setting       & $\langle p_{t} \rm{kick} \rangle $ [GeV/c]         & (\rm{g/cm$^3$})   & (\rm{g/cm$^3$}) \\ \hline
 low	       &     -1.04 / ~4.72     & $ 0.07066$  & $0.16280$ \\ \hline
 intermediate  &     ~~~0~ / ~4.72       & $ 0.07062$  & $0.16272$ \\
 	           & ~~~0~ / -4.72       & $ 0.07064$  & $0.16280$ \\ \hline
  	           & ~~~0~ / ~6.39       & $ 0.07064$  & $0.16278$ \\
 high  	       &     ~~~0~ / ~6.39       & $ 0.07062$  & $0.16265$ \\
               &     ~~~0~ / -6.39       & $ 0.07061$  & $0.16259$ \\
\end{tabular}
\end{center}
\end{table}

The beam was attenuated as it interacted with the target material.  
Since the deuterium target had the higher density  
the beam intensity decreased
more rapidly as it passed through the deuterium target.
Calculations based on the proton-proton and proton-deuteron cross sections~\cite{atten1,atten2,atten3,atten4}
were used to determine the ratio of the effective luminosity in the
hydrogen target, $A_h$, to the effective luminosity in the deuterium target, $A_{d}$:
\begin{equation}
\frac{A_h}{A_d} = 1.042 \pm 0.002.
\end{equation}

The acceptances for the events from the hydrogen and deuterium 
targets were not identical.  Although the target-flask construction 
and location were identical, the attenuation of the beam through the targets 
meant that the average interaction points for the two targets were 
slightly different.  The average interaction point in the deuterium
target was $\approx 0.5$~\rm{cm} upstream of that for the hydrogen target.  
Monte Carlo simulations were done to study the effects 
of beam attenuation on the acceptance. 
These studies gave a slight $x_2$-dependent correction. 
The maximum size of this correction 
was about 1\% at the highest $x_2$ data points in
the low and intermediate mass data.  The typical correction
was an order of magnitude smaller.

\section{Calculation of $\sigma^{\lowercase{pd}}/2\sigma^{\lowercase{pp}}$}

This experiment counted the number of dimuon events, $N$, from the 
hydrogen, deuterium, and empty targets.  
To compare the yields from these
targets, the beam intensity for each spill was recorded and the 
integrated beam intensity, $I$, for each target 
was determined.  Using the many small corrections previously 
described, the number of raw hydrogen dimuon events is
\begin{equation}
N_h = I_h A_h t_h \rho_h \left[ \frac{H}{g} \right] \frac{d\sigma^{pp}}{d\Omega} 
      \Delta\Omega_h e_h + N_h^{\rm BG},
\end{equation}
and the number of raw deuterium events is
\begin{equation}                                                      
N_d = I_d A_d t_d \rho_d \left[ \frac{D}{g} \right] \frac{d\sigma^{pd}}{d\Omega}     
      \Delta\Omega_d e_d + N_d^{\rm BG}.                                    
\label{eqn:ndeter}
\end{equation}                                                         
In the equations in this section, the subscripts indicate the target 
type, hydrogen, $h$, deuterium, $d$, and empty, $e$.
The target length is $t$, 
$H/g$ and $D/g$ are the number of hydrogen and deuterium atoms per gram,
$\rho$ is the target density,
$\Delta\Omega$ is the spectrometer acceptance for a 
given target, $e$ is the detector efficiency for a given target, 
and $N{\rm^{BG}}$ is the number of background 
events for a given target.
Using these equations, one obtains 
\begin{equation}
\frac{\sigma^{pd}}{2\sigma^{pp}} =
\frac{1}{2}
\frac{N_d - N_d^{\rm BG}}{N_h - N_h^{\rm BG}}~
\left[ \frac{I_h}{I_d}
\frac{A_h}{A_d}
\frac{t_h}{t_d}
\frac{\rho_h}{\rho_d}
\frac{H/g}{D/g}
\frac{\Delta\Omega_h}{\Delta\Omega_d} 
\frac{e_h}{e_d} \right].
\end{equation}
Note that the quantity in brackets is $\approx 1$.

The small amount of hydrogen contamination in the deuterium target after
it was filled the second time was accounted for by altering  
Eq. \ref{eqn:ndeter} to read

\begin{equation}                                                      
N_d = I_d A_d t_d \rho^{\,\prime}_d \frac{D}{g} 
\bigl(f_d \frac{d\sigma^{pd}}{d\Omega}+
f_h \frac{d\sigma^{pp}}{d\Omega} \bigr)
\Delta\Omega_d e_d + N_d^{\rm BG}.
\end{equation}
In the equation above, $f_d$ and $f_h$ are the percent by volume of
deuterium and hydrogen respectively in the deuterium target, and
$\rho^{\,\prime}_d$ is the density of the contaminated deuterium. The ratio of
Drell-Yan cross sections is then
\begin{eqnarray}
\lefteqn{\frac{\sigma^{pd}}{2\sigma^{pp}} =} \hspace{0.39in} \\
\nonumber
& &
\frac{1}{2}
\frac{N_d - N_d^{\rm BG}}{N_h - N_h^{\rm BG}}
\left[ \frac{I_h}{I_d}
\frac{A_h}{A_d}
\frac{t_h}{t_d}
\frac{\rho_h}{f_d \rho^{\,\prime}_d}
\frac{H/g}{D/g}
\frac{\Delta\Omega_h}{\Delta\Omega_d}
\frac{e_h}{e_d} \right]
 - \frac{f_h}{2f_d}.
\end{eqnarray}

The background events originated from two separate production
mechanisms.  The first source was Drell-Yan events produced from
beam interactions with the target flask windows or other non-target materials.
The number of these events was determined by normalizing the yields from
the empty target.  To properly normalize the number of 
empty-target events from downstream of the center of the target, attenuation 
of the beam through the target must be included.
The second source of background events was the
randoms ($N_{\rm {target}}^{\rm {randoms}}$) 
that were described previously.
Combining these two sources gives
\begin{equation}
N_h^{\rm BG} = \left(N_e^{\rm up} + 0.93N_e^{\rm down}\right)
	   \frac{I_h}{I_e} + N_h^{\rm randoms}
\end{equation}
for the hydrogen target background and 
\begin{equation}
N_d^{\rm BG} = \left(N_e^{\rm up} + 0.85N_e^{\rm down}\right)
	   \frac{I_d}{I_e} + N_d^{\rm randoms}
\end{equation}
for the deuterium target background.  
In the previous two equations the superscript on $N_e$ designates 
whether the empty target event originated from upstream or downstream
of the center of the target.  
Typical empty target corrections are 12\% for hydrogen and 5\% for deuterium.

The output of the second-pass analysis was subjected to the 
quality cuts described earlier. Events that passed the cuts, after
being corrected for random and non-target events as described above, were
used to determine $\sigma^{pd}/2\sigma^{pp}$ versus $x_2$.  
These results are shown in
Tables \ref{tab:hidhigh}, \ref{tab:hidint}, and \ref{tab:hidlow}.
The results shown for the high-mass data are slightly different 
from and supersede those previously published~\cite{prl}, due to
minor improvements made to the rate dependence and acceptance calculations.  
The average values of $x_2$, $x_F$, $p_T$, and dimuon mass are also shown
in Tables \ref{tab:hidhigh}, \ref{tab:hidint}, and \ref{tab:hidlow}.

\begin{figure}
  \begin{center}
    \mbox{\epsffile{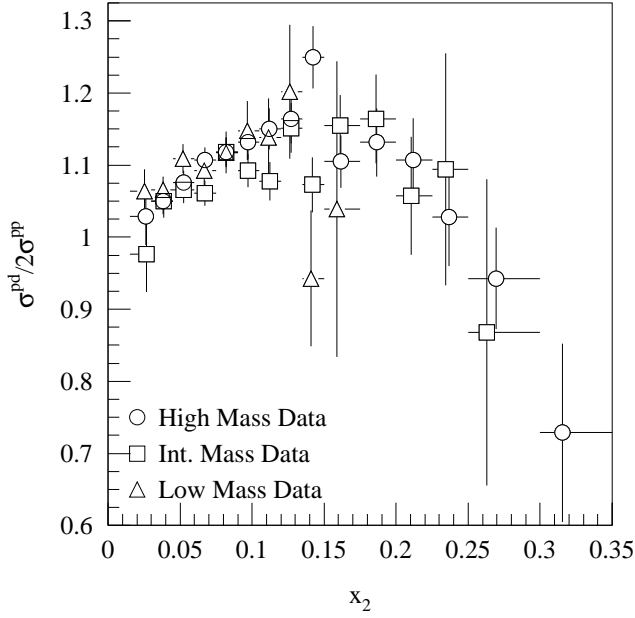}}
    \vspace*{-0.05in}                                
  \end{center}       
  \caption{The Drell-Yan cross section ratio versus $x_{2}$.
	   The results from all three mass settings are shown.  The error bars
	   represent the statistical uncertainty.  The systematic uncertainty 
	   is less than 1\% for each data set as shown in Table~\ref{tab:systematics}.}
  \label{fig:ratio}                    
\end{figure}

\begin{table}[b!]
  \caption{Cross section ratios binned in $x_2$, with their statistical 
  uncertainties and average values for kinematic
          variables for the high mass data. Systematic uncertainties are
          reported in Table \ref{tab:systematics}.
}
  \label{tab:hidhigh}
  \begin{center}                           
    \begin{tabular}{c|ccccc}
$x_2$ range &         &         & $\langle p_T\rangle$ & $\langle M_{\mu^+\mu^-}\rangle$ \\
  min-max   & $\langle x_2\rangle$ & $\langle x_F\rangle$ & (\rm{GeV/c}) &    (\rm{GeV/c$^2$})  
& $\sigma^{pd}/2\sigma^{pp}$ \\ \hline
0.015-0.030 & 0.026 & 0.624 & 0.842 &  5.0 & 1.029 $\pm$ 0.040 \\
0.030-0.045 & 0.038 & 0.520 & 0.935 &  5.6 & 1.050 $\pm$ 0.018 \\
0.045-0.060 & 0.053 & 0.456 & 1.009 &  6.3 & 1.075 $\pm$ 0.016 \\
0.060-0.075 & 0.067 & 0.411 & 1.085 &  6.9 & 1.107 $\pm$ 0.018 \\
0.075-0.090 & 0.082 & 0.367 & 1.133 &  7.4 & 1.118 $\pm$ 0.020 \\
0.090-0.105 & 0.097 & 0.319 & 1.168 &  7.8 & 1.131 $\pm$ 0.023 \\
0.105-0.120 & 0.112 & 0.279 & 1.185 &  8.1 & 1.150 $\pm$ 0.029 \\
0.120-0.135 & 0.127 & 0.250 & 1.202 &  8.4 & 1.164 $\pm$ 0.034 \\
0.135-0.150 & 0.142 & 0.230 & 1.209 &  8.8 & 1.249 $\pm$ 0.043 \\
0.150-0.175 & 0.162 & 0.213 & 1.211 &  9.4 & 1.105 $\pm$ 0.036 \\
0.175-0.200 & 0.186 & 0.185 & 1.206 & 10.0 & 1.132 $\pm$ 0.047 \\
0.200-0.225 & 0.212 & 0.160 & 1.173 & 10.7 & 1.107 $\pm$ 0.057 \\
0.225-0.250 & 0.237 & 0.128 & 1.201 & 11.2 & 1.028 $\pm$ 0.069 \\
0.250-0.300 & 0.269 & 0.093 & 1.180 & 12.0 & 0.943 $\pm$ 0.071 \\
0.300-0.350 & 0.315 & 0.046 & 1.078 & 12.9 & 0.729 $\pm$ 0.124 \\
    \end{tabular}
  \end{center}
\end{table}

\begin{table}
  \caption{Cross section ratios binned in $x_2$, with their statistical 
  uncertainties and average values for kinematic
          variables for the intermediate mass data.   Systematic uncertainties are
          reported in Table \ref{tab:systematics}.}
  \label{tab:hidint}
  \begin{center}                           
    \begin{tabular}{c|ccccc}
$x_2$ range &         &         & $\langle p_T\rangle$ & $\langle M_{\mu^+\mu^-}\rangle$ \\
  min-max   & $\langle x_2\rangle$ & $\langle x_F\rangle$ & (\rm{GeV/c}) &    
  (\rm{GeV/c$^2$})  
& $\sigma^{pd}/2\sigma^{pp}$\\ \hline
0.015-0.030 & 0.027 & 0.514 & 1.296 &  4.6 & 0.976 $\pm$ 0.052 \\
0.030-0.045 & 0.039 & 0.386 & 1.179 &  4.9 & 1.050 $\pm$ 0.023 \\
0.045-0.060 & 0.053 & 0.329 & 1.152 &  5.4 & 1.065 $\pm$ 0.018 \\
0.060-0.075 & 0.067 & 0.297 & 1.142 &  6.0 & 1.061 $\pm$ 0.018 \\
0.075-0.090 & 0.082 & 0.265 & 1.140 &  6.5 & 1.118 $\pm$ 0.021 \\
0.090-0.105 & 0.097 & 0.230 & 1.144 &  6.9 & 1.092 $\pm$ 0.023 \\
0.105-0.120 & 0.112 & 0.195 & 1.160 &  7.1 & 1.078 $\pm$ 0.027 \\
0.120-0.135 & 0.127 & 0.161 & 1.154 &  7.4 & 1.152 $\pm$ 0.035 \\
0.135-0.150 & 0.142 & 0.134 & 1.118 &  7.6 & 1.073 $\pm$ 0.038 \\
0.150-0.175 & 0.161 & 0.107 & 1.095 &  7.9 & 1.155 $\pm$ 0.042 \\
0.175-0.200 & 0.186 & 0.081 & 1.045 &  8.4 & 1.164 $\pm$ 0.062 \\
0.200-0.225 & 0.211 & 0.070 & 1.080 &  9.2 & 1.057 $\pm$ 0.082 \\
0.225-0.250 & 0.234 & 0.079 & 1.055 & 10.3 & 1.094 $\pm$ 0.161 \\
0.250-0.300 & 0.263 & 0.153 & 1.135 & 12.7 & 0.868 $\pm$ 0.213 \\
    \end{tabular}
  \end{center}
\end{table}

\begin{table}
  \caption{Cross section ratios binned in $x_2$, with their statistical 
  uncertainties and average values for kinematic
          variables for the low mass data.  Systematic uncertainties are
          reported in Table \ref{tab:systematics}.} 
  \label{tab:hidlow}
  \begin{center}                           
    \begin{tabular}{c|ccccc}
$x_2$ range &         &         & $\langle p_T\rangle$ & $\langle M_{\mu^+\mu^-}\rangle$ \\
  min-max   & $\langle x_2\rangle$ & $\langle x_F\rangle$ & 
  (\rm{GeV/c}) &    (\rm{GeV/c$^2$})  
& $\sigma^{pd}/2\sigma^{pp}$ \\ \hline
0.015-0.030 & 0.025 & 0.495 & 0.992 &  4.4 & 1.064 $\pm$ 0.030 \\
0.030-0.045 & 0.038 & 0.351 & 1.036 &  4.7 & 1.066 $\pm$ 0.018 \\
0.045-0.060 & 0.052 & 0.275 & 1.069 &  5.0 & 1.109 $\pm$ 0.020 \\
0.060-0.075 & 0.067 & 0.238 & 1.076 &  5.5 & 1.092 $\pm$ 0.023 \\
0.075-0.090 & 0.082 & 0.210 & 1.065 &  5.9 & 1.118 $\pm$ 0.029 \\
0.090-0.105 & 0.097 & 0.182 & 1.057 &  6.3 & 1.148 $\pm$ 0.041 \\
0.105-0.120 & 0.112 & 0.151 & 1.035 &  6.6 & 1.138 $\pm$ 0.055 \\
0.120-0.135 & 0.126 & 0.129 & 1.051 &  6.9 & 1.202 $\pm$ 0.093 \\
0.135-0.150 & 0.141 & 0.118 & 1.055 &  7.4 & 0.943 $\pm$ 0.094 \\
0.150-0.175 & 0.159 & 0.091 & 1.007 &  7.7 & 1.039 $\pm$ 0.205 \\
    \end{tabular}
  \end{center}
\end{table}

\begin{table}
\caption{Systematic uncertainties in the measurement of $\sigma^{pd}/2\sigma^{pp}$.}
\label{tab:systematics}
\begin{center}
\begin{tabular}{cccc}
\multicolumn{1}{c}{source of} &
\multicolumn{3}{c}{mass setting}  \\
 uncertainty  	& high		& intermediate	& low  \\
\hline
rate dependence	& 0.69\%	& 0.89\%	& 0.82\% \\
target length	& 0.2~\%	& 0.2~\% 	& 0.2~\% \\
beam intensity 	& 0.1~\%	& 0.1~\%	& 0.1~\% \\
attenuation/acceptance & 0.2~\% & 0.2~\% 	& 0.2~\% \\
deuterium composition & 0.61\%	&  ---		& ---     \\
\hline
total           & 0.97\%       & 0.94\%       & 0.87\% \\
\end{tabular}
\end{center}
\end{table}

\begin{figure}
  \begin{center}
    \mbox{\epsfxsize=8.5cm\epsffile{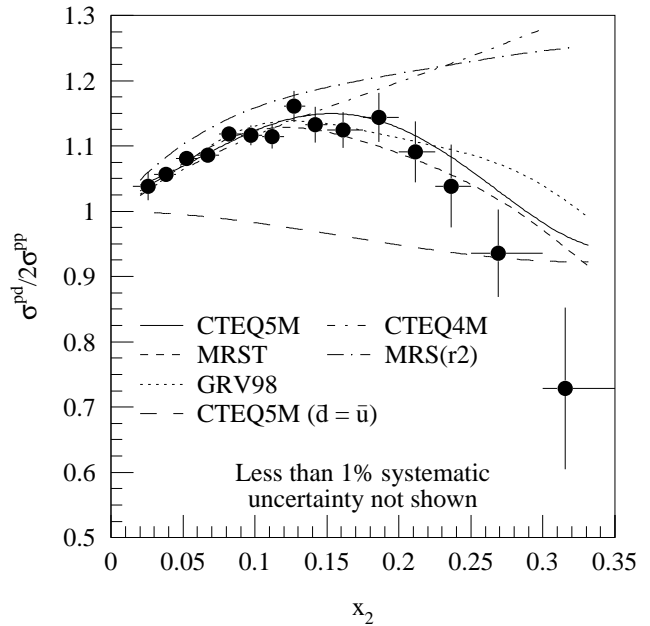}}                      
    \vspace*{-0.05in}                                                      
  \end{center}                                                               
  \caption{The Drell-Yan cross section ratio versus $x_{2}$.
           The results from all three mass settings have been combined.  
	   The error bars represent the statistical uncertainty.  
	   The systematic uncertainty is common to all points and is less 
	   than 1\%. The curves 
	   are  the calculated next-to-leading-order cross-section ratios using 
	   various parton distribution functions.  The bottom curve is calculated
	   using CTEQ5M where $\bar d - \bar u$ has been forced to zero.}
	   \label{fig:ratiowpdf}
\end{figure}

The average 
cross-section ratios for each mass setting are shown
in Fig. \ref{fig:ratio}. The three mass settings agree
and are consistent within their systematic uncertainties. 
The result of averaging all of 
the mass settings is shown in Fig. \ref{fig:ratiowpdf}
and Table \ref{tab:hidall}.

Since this is a measurement of cross-section ratios, 
the only sources of systematic uncertainty that must be 
considered are those that affect the two targets differently.   
Because the targets were changed every few minutes, effects such
as changes in detector efficiency or beam quality were minimized.

The important sources of systematic uncertainty include
differences in the
rate dependence, target flask length, target composition, beam attenuation,
and acceptance.  Table~\ref{tab:systematics} shows the main 
sources of systematic uncertainty in the cross section ratio 
for each mass setting.  Clearly the rate dependence and deuterium composition
are the dominant uncertainties. Adding all of the 
sources of systematic uncertainties in quadrature, the 
total systematic uncertainty in the measured cross section
ratio is less than 1\%.

\begin{table*}
  \caption{The cross section ratio, $\bar d / \bar u$ and $\bar d - \bar u$
  values determined from the combination of all data sets for each
  $x_2$ bin.  The first uncertainty is statistical, the second
  systematic.  The cross section ratio has a systematic uncertainty of
  less than 1\% as shown in Table \ref{tab:systematics}.  
  The average values for kinematic variables are also shown.}
  \label{tab:hidall}
  \begin{center}
    \begin{tabular}{c|ccccccc}
$x_2$ range &         &         & $\langle p_T\rangle$ & $\langle M_{\mu^+\mu^-}\rangle$ & & & \\
  min-max   & $\langle x_2\rangle$ & $\langle x_F\rangle$ & (\rm{GeV/c}) &    
  (\rm{GeV/c$^2$})  &
$\sigma^{pd}/2\sigma^{pp}$ & $\bar{d}/\bar{u}$ & $ \bar{d} - \bar{u}$ \\ \hline
0.015-0.030 & 0.026 & 0.534 & 1.004 &  4.6 & 1.038 $\pm$ 0.022 & 1.085 $\pm$ 0.050 $\pm$ 0.017 &  0.862 $\pm$ 0.489 $\pm$ 0.167 \\
0.030-0.045 & 0.038 & 0.415 & 1.045 &  5.1 & 1.056 $\pm$ 0.011 & 1.140 $\pm$ 0.027 $\pm$ 0.018 &  0.779 $\pm$ 0.142 $\pm$ 0.096 \\
0.045-0.060 & 0.052 & 0.356 & 1.076 &  5.6 & 1.081 $\pm$ 0.010 & 1.215 $\pm$ 0.026 $\pm$ 0.020 &  0.711 $\pm$ 0.077 $\pm$ 0.060 \\
0.060-0.075 & 0.067 & 0.326 & 1.103 &  6.2 & 1.086 $\pm$ 0.011 & 1.249 $\pm$ 0.028 $\pm$ 0.021 &  0.538 $\pm$ 0.055 $\pm$ 0.041 \\
0.075-0.090 & 0.082 & 0.296 & 1.122 &  6.8 & 1.118 $\pm$ 0.013 & 1.355 $\pm$ 0.036 $\pm$ 0.023 &  0.512 $\pm$ 0.044 $\pm$ 0.028 \\
0.090-0.105 & 0.097 & 0.261 & 1.141 &  7.2 & 1.116 $\pm$ 0.015 & 1.385 $\pm$ 0.046 $\pm$ 0.025 &  0.400 $\pm$ 0.040 $\pm$ 0.022 \\
0.105-0.120 & 0.112 & 0.227 & 1.156 &  7.5 & 1.115 $\pm$ 0.018 & 1.419 $\pm$ 0.060 $\pm$ 0.027 &  0.321 $\pm$ 0.038 $\pm$ 0.017 \\
0.120-0.135 & 0.127 & 0.199 & 1.168 &  7.8 & 1.161 $\pm$ 0.023 & 1.630 $\pm$ 0.085 $\pm$ 0.031 &  0.338 $\pm$ 0.034 $\pm$ 0.013 \\
0.135-0.150 & 0.142 & 0.182 & 1.161 &  8.2 & 1.132 $\pm$ 0.027 & 1.625 $\pm$ 0.110 $\pm$ 0.033 &  0.259 $\pm$ 0.035 $\pm$ 0.010 \\
0.150-0.175 & 0.161 & 0.164 & 1.156 &  8.7 & 1.124 $\pm$ 0.027 & 1.585 $\pm$ 0.111 $\pm$ 0.032 &  0.180 $\pm$ 0.027 $\pm$ 0.008 \\
0.175-0.200 & 0.186 & 0.146 & 1.146 &  9.5 & 1.144 $\pm$ 0.038 & 1.709 $\pm$ 0.158 $\pm$ 0.036 &  0.142 $\pm$ 0.023 $\pm$ 0.005 \\
0.200-0.225 & 0.211 & 0.133 & 1.146 & 10.3 & 1.091 $\pm$ 0.047 & 1.560 $\pm$ 0.194 $\pm$ 0.034 &  0.081 $\pm$ 0.022 $\pm$ 0.004 \\
0.225-0.250 & 0.236 & 0.120 & 1.178 & 11.1 & 1.039 $\pm$ 0.063 & 1.419 $\pm$ 0.264 $\pm$ 0.036 &  0.045 $\pm$ 0.023 $\pm$ 0.003 \\
0.250-0.300 & 0.269 & 0.097 & 1.177 & 12.0 & 0.935 $\pm$ 0.067 & 1.082 $\pm$ 0.256 $\pm$ 0.032 &  0.006 $\pm$ 0.019 $\pm$ 0.002 \\
0.300-0.350 & 0.315 & 0.046 & 1.078 & 12.9 & 0.729 $\pm$ 0.124 & 0.346 $\pm$ 0.395 $\pm$ 0.022 & -0.040 $\pm$ 0.036 $\pm$ 0.002 \\
    \end{tabular}
  \end{center}
\end{table*}

    \section{Extraction of \lowercase{$\bar d(x)/\bar u(x)$}}

    From the discussion in Section I, it is clear that $\sigma^{pd}/2\sigma^{pp}$ is
    closely related to $\bar d/\bar u$. However, the simple
    approximations that lead to Eq. \ref{eqn:prl41} are not fully satisfied
    since the data cover a larger range in $x_{F}$. Therefore, an
    iterative process was used to extract $\bar d/\bar u$ versus $x_2$ from the 
    cross-section ratio.

\begin{figure}

    \begin{center}
    \mbox{\epsffile{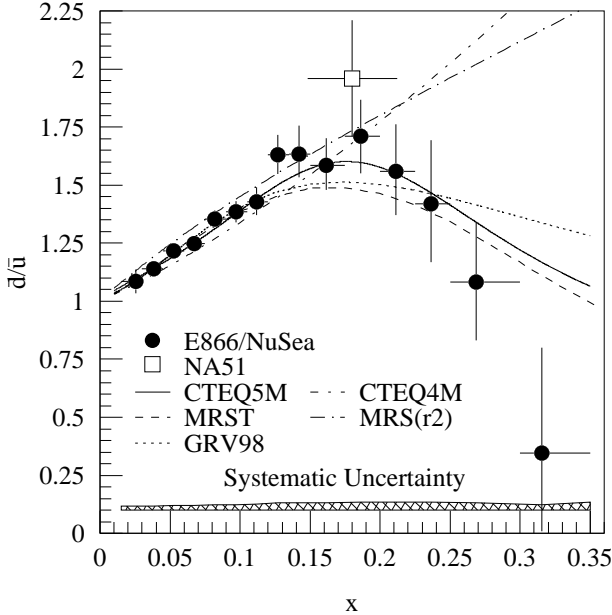}}
    \vspace*{-0.05in}
    \end{center}

    \caption{$\bar d(x)/\bar u(x)$ versus $x$ shown with statistical and
    systematic uncertainties. The combined result from all three mass settings
    is shown with various parameterizations at $Q^2=54.0$~GeV$^2/$c$^2$. The NA51
    data point is also shown.}

    \label{fig:duwpdf}
    \end{figure}

    The iterative process calculated $\sigma^{pd}/2\sigma^{pp}$ at
    leading order,\footnote{The
    difference between next-to-leading-order and leading-order
    calculations of the cross section {\em ratio} in the region of
    interest is less than 2.1\%.} 
    folded it with the experimental acceptance, and compared this calculated quantity
    with the measurement.  Next, the $\bar d/\bar u$ ratio 
    was adjusted to improve
    the agreement. This process continued until the calculated
    $\sigma^{pd}/2\sigma^{pp}$ agreed with the measured ratio. The
    results of this method, using the combined data from all mass
    settings, are shown in Fig. \ref{fig:duwpdf} together with parameterizations 
    from various PDFs \cite{CTEQ,MRS,CTEQ5,mrst,grv98}.

    It is clear from Eq. 4 that the calculation of
    $\sigma^{pd}/2\sigma^{pp}$ requires the PDF for each quark and
    antiquark in the proton as input. 
    In the iterative process, it was assumed
    that existing PDF parameterizations accurately describe the valence
    and heavy-quark distributions as well as the quantity $\bar d(x) +
    \bar u(x)$, since these quantities have been constrained by previous
    measurements.  The parameterizations used
    were CTEQ5M~\cite{CTEQ5} and MRST~\cite{mrst}.

    For the calculated $\sigma^{pd}/2\sigma^{pp}$ to be compared to
    the measured $\sigma^{pd}/2\sigma^{pp}$, the acceptance of the
    spectrometer must be included in the calculation. To do this the
    cross section ratio was calculated for the $x_1$, $x_2$, and $Q^2$
    values of every event that passed the analysis cuts. These calculated
    cross section ratios were then averaged over each $x_2$ bin.

    As $\sigma^{pd}/2\sigma^{pp}$ was calculated for each iteration, it
    was assumed that $\bar d/\bar u$ for the beam proton was the same as
    $\bar d/\bar u$ for the target proton over the $x_2$ range of the
    data. For many events however, $x_1$ was greater than the maximum
    $x_2$ in the data, so some assumption was required for the value of $\bar
    d(x_1)/\bar u(x_1)$ for $x_1 \geq 0.35$. The effects of several
    different assumptions were investigated. The extracted
    $\bar d/\bar u$ was not noticeably affected by any of these
    assumptions except at the highest $x$ values, where $\bar d/\bar u$ was affected by less
    than five percent. The assumption finally used was $\bar d(x_1)/\bar
    u(x_1) \equiv 1.0$ in the proton for $x_1 > 0.35$.

\begin{figure}
    \begin{center}
    \mbox{\epsffile{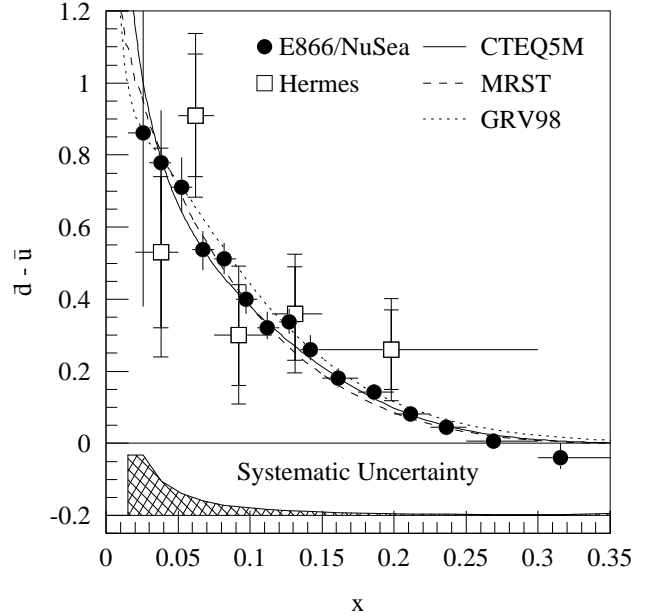}}
    \vspace*{-0.05in}
    \end{center}

    \caption{$\bar d - \bar u$ as a function of $x$ shown with
    statistical and systematic uncertainties. The E866 results, scaled to
    fixed $Q^2 = 54$~GeV$^2/$c$^2$, are shown as the circles. Results from HERMES
    ($\langle Q^2\rangle = 2.3$~GeV$^2/$c$^2$) are shown as squares. The error bars on the E866 data
    points represent the statistical uncertainty. The inner error bars
    on the HERMES data points represent the statistical uncertainty
    while the outer error bars represent the statistical and systematic
    uncertainty added in quadrature.}

    \label{fig:dmu}
    \end{figure}

\begin{figure}
    \begin{center}
    \mbox{\epsffile{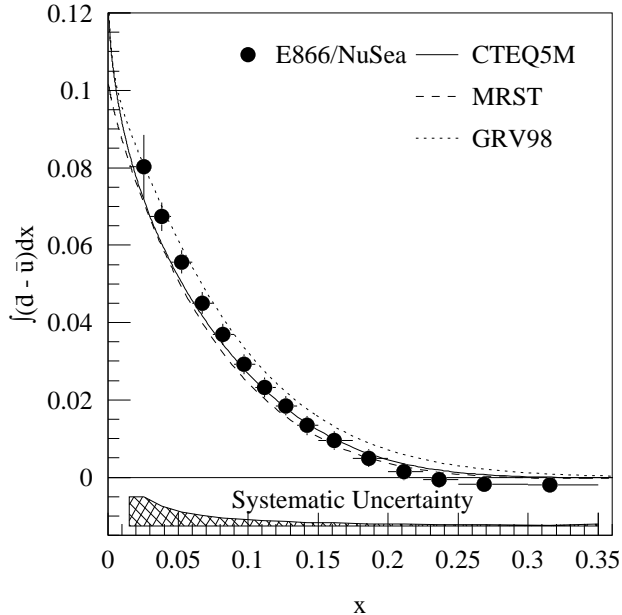}}
    \vspace*{-0.05in}
    \end{center}

    \caption{$\int^{0.35}_x \left[\bar d(x') - \bar u(x')\right] dx'$
    versus $x$ shown with statistical and systematic uncertainties at fixed $Q^2
    = 54$~GeV$^2/$c$^2$. The curves are from three different
    parameterizations.}

    \label{fig:idmu}
    \end{figure}

    Once the quantity $\bar d(x)/\bar u(x)$ was determined, the quantity
    $\bar d(x)-\bar u(x)$ was calculated, again assuming that the quantity
    $\bar d(x) + \bar u(x)$ is well described by the parameterizations.
    So that $\bar d(x)-\bar u(x)$ could be integrated, the $\bar d(x)/\bar u(x)$
    values were scaled to a fixed $Q^2$, with $Q = 7.35$~GeV/c. The scaling
    procedure multiplied $\bar d(x,Q)/\bar u(x,Q)$ by the ratio
    $\frac{\bar d(x,Q=7.35)/\bar u(x,Q=7.35)}{\bar d(x,Q)/\bar u(x,Q)}$ as
    give by CTEQ5M.  (There was no significant difference if MRST was used instead of CTEQ5M.)
    Figure \ref{fig:dmu} and Table \ref{tab:hidall}
    show $\bar d(x)-\bar u(x)$ as a function of $x$. These data can be
    integrated over $x$ to provide $\int^1_0 \left[\bar d(x) - \bar
    u(x)\right] dx = 0.118 \pm 0.012$ for the proton.  An extrapolation was made to account for the
    unmeasured region at low $x$. 
To extrapolate this integral from the measured region, which is shown in
Fig. \ref{fig:idmu}, to the unmeasured 
region, MRST and CTEQ5M were used to estimate the contribution for $0 \le x \le 0.015$ and
it was assumed that the contribution for $x \ge 0.35$ was negligible.  
The uncertainty from this extrapolation was estimated to be 0.0041 which is half the difference 
between the contributions as given by MRST and CTEQ5M.

\section{Charge Symmetry and Shadowing}

 The analysis presented here assumes that the parton distributions of the
    nucleon obey charge symmetry: {\it i.e.}, $u_p(x) = d_n(x)$,
    $\bar{d}_p(x) = \bar{u}_n(x)$, {\it etc}. This is consistent with the
    treatment in previous experiments \cite{prl,nmc,na51,hermes} and
    global fits \cite{CTEQ,MRS,grv95}. The possibility that charge
    symmetry could be significantly violated (CSV) at the parton level has been
    discussed by several authors 
    \cite{ma1,ma2,sather,rodionov,benesh1,benesh2,londergan2}
    and an extensive review was recently published \cite{londergan2}.

    Using the cloudy-bag model, it has been demonstrated \cite{rodionov}
    that a CSV effect of $\approx\,5\%$ could exist for the
    ``minority valence quarks" [{\it i.e.}, $d_p(x)$ and $u_n(x)$] at $x>0.4$. 
    In contrast, a study \cite{benesh2} of sea quark CSV showed
    it to be roughly a factor of 10 smaller than CSV for valence quarks. This
    was called into question in an analysis by Boros {\it et al}. 
    \cite{boros1,boros2} of the $F_2$ structure functions determined from muon
    and neutrino deep inelastic scattering, which concluded that $\bar{d}_n(x)
    \approx 1.25 \bar{u}_p(x)$ at small $x$. However, Bodek {\it et al}.
    \cite{bodek99} showed that $W$ charge
    asymmetry measurements are inconsistent with the CSV effect identified by
    Boros {\it et al}. and consistent with the assumption of sea quark charge
    symmetry. Subsequently, a more recent work by Boros {\it et al}.
    \cite{boros3} concluded that, after corrections are made for
    nuclear shadowing in the neutrino-induced data and the charm production
    threshold is treated explicitly using NLO QCD, the deep inelastic muon and
    neutrino scattering data provide no evidence for sea quark CSV.

    Throughout the above analysis, we have assumed that nuclear effects
    in deuterium may be neglected, so that $\sigma^{pd} = \sigma^{pp} +
    \sigma^{pn}$. This is consistent with the traditional approach, in which
    nuclear effects in deuterium are included in global parton distribution fits
    \cite{CTEQ5,mrst,grv98} and neglected in experimental analyses
    \cite{prl,nmc,na51,hermes}. However, it is important to estimate the
    magnitude of these corrections. The nuclear dependence of proton-induced 
    Drell-Yan dimuon production at 800 GeV/c has been
    determined by FNAL E866/NuSea \cite{Vasi99} and by FNAL
    E772 \cite{Alde90}. These experiments measured the relative
    Drell-Yan cross sections per nucleon on a range of nuclear targets. Both
    experiments find little, if any, nuclear dependence for $x>0.08$. In this
    region, we may conservatively estimate that any nuclear effects in
    the proton-deuterium Drell-Yan cross section are $<\,0.5$\%. However, at small $x$, the nuclear data show clear
    evidence for nuclear shadowing. In principle, one may use the
    parameterization $\sigma^{pA} = \sigma_0 A^{\alpha}$, where A is 
    the atomic number, to extrapolate the
    observed effects in heavier nuclei to deuterium. But this will overestimate
    them, due to the anomalously large internucleon separation in the
    deuteron. 
    
    Alternatively, one may note that the shadowing effects seen in Drell-Yan
    by E866 \cite{Vasi99} and in deep inelastic
    scattering by NMC \cite{NMC_Adep} are nearly equal, in spite of
    the different reaction mechanisms and momentum transfers of the two
    experiments, so we may use calculations of shadowing in deep inelastic
    scattering \cite{shadow,Bade94} to set the scale of the nuclear
    effects that may be present in our deuterium data. We estimate that
    shadowing implies a reduction of 0.9\% to $\sigma^{pd}$ in Eq. \ref{eqn:depn}
    for our smallest $x_2$ point, based on the calculations of Badelek and
    Kwiecinski \cite{Bade94}. This would increase
    $\bar{d}(x)/\bar{u}(x)$ by $<\,2$\% in our $x$ range. Our extracted value of
    $(\bar{d}-\bar{u})|_{x=0.026}$ would increase by 26\%. The correction
    to $\bar{d}-\bar{u}$ drops very rapidly as $x$ increases. Our value for
    $\int_{0.015}^{0.35}(\bar{d}-\bar{u})dx$ would increase by 10\%.
    The nuclear effects in deuterium, and hence the corrections to our
    results, are estimated to be approximately half as large in the calculations
    of Melnitchouk and Thomas \cite{shadow}. We conclude that the
    correction due to shadowing in deuterium may be comparable to our
    systematic uncertainty for our smallest $x$ values, and is much smaller
    than our systematic uncertainty for $x>0.06$.

\section{Dependence on Other Kinematic Variables}

\begin{figure}       
  \begin{center}
    \mbox{\epsfxsize=8.5cm\epsffile{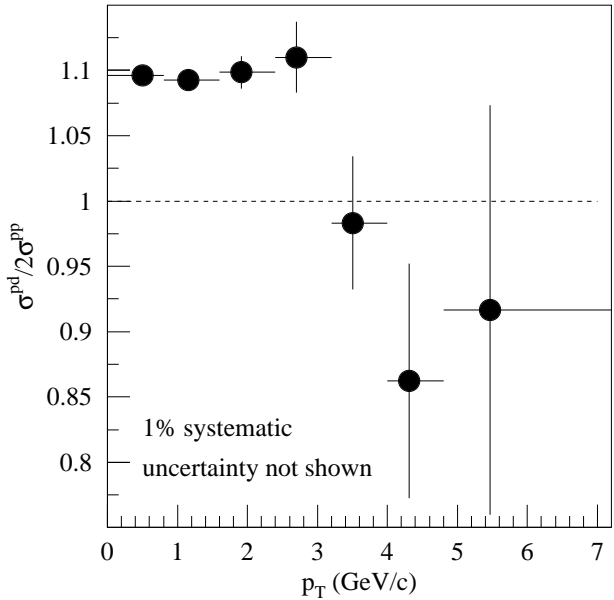}}
  \end{center}                                                                     
  \caption{The Drell-Yan cross section ratio versus $p_T$.
           The combined result from all data sets is shown.  The error bars     
           represent the statistical uncertainty.  There is a one percent
           systematic uncertainty common to all points.}
   \label{fig:ratiopt}                                                               
\end{figure}

The cross section ratio for deuterium versus hydrogen can be studied
as a function of kinematic quantities other than $x_2$ .  
Figure \ref{fig:ratiopt} shows the ratio as a function of the 
transverse momentum of the dimuon. Studies of the data and of
Monte Carlo acceptance calculations show that the observed shape
versus $p_T$ is not due to acceptance differences between the
targets or correlations with $x_2$.

For $p_T$ values below 3 \rm{GeV/c} there may be evidence for a slight rise in the
ratio with $p_T$, consistent with a small amount of additional
multiple scattering of the incoming parton in deuterium. 
Above 3~\rm{GeV/c} the ratio drops abruptly to near or below unity.
This could be a signature for a change in reaction mechanism. 

Recently, Berger {\it et al}.\@ \cite{Berger98} calculated the $p_T$
dependence of the 
Drell-Yan cross section off an (isoscalar) nucleon to $O({\alpha_s}^2)$,
including the modifications at small $p_T$ due to 
all-orders soft-gluon resummation.  They find that the 
quark-antiquark annihilation process $q{\bar q} \rightarrow \gamma^* X$
dominates the 
Drell-Yan yield at small $p_T$, and the 
quark-gluon Compton scattering process $qg \rightarrow q\gamma^* X$
dominates at large $p_T$.  This implies that the sensitivity of
$\sigma^{pd}/2\sigma^{pp}$ to ${\bar d}/{\bar u}$ arises primarily at small
$p_T$, while the 
large-$p_T$ ratio measures the relative gluon densities in the proton and
deuteron.  The calculations indicate that the crossover between the two
processes occurs at $p_T \sim$ 2 to 3~\rm{GeV/c} for the kinematics of the E866
data, close to the point where the 
cross-section ratio versus $p_T$ in Fig. \ref{fig:ratiopt} begins to drop.  Thus, the
E866 $\sigma^{pd}/2\sigma^{pp}$ results may also provide information
regarding the gluonic composition of the nucleon, but such an analysis is
outside the scope of the present paper.

The dependences of the deuterium to hydrogen ratio on $x_1$,  $x_F$, and
dimuon mass were also studied.  Unlike $p_T$, these studies showed no independent 
dependence on these kinematic variables, reinforcing the 
conclusion that $x_2$ is the important variable for our data.

\section{Comparison to Other Results}

The results of this experiment are much more extensive
and precise than any other measurement of $\bar d(x)/\bar u(x)$.
Other measurements of $\bar d(x)/\bar u(x)$ include the early 
measurement by NA51 and
the recent result from the HERMES collaboration at
DESY.  These measurements are in general agreement with 
the E866 results as seen in Fig.~\ref{fig:duwpdf} and Fig.~\ref{fig:dmu}. 
Even though the average $Q^2$ values of these measurements differ,
comparisons can be made between them because the $Q^2$ dependence 
is small.

While the NA51 determination of $\bar d(x)/\bar u(x)$ was very
similar to the method used by E866, the HERMES result was 
based on a measurement of semi-inclusive deep-inelastic scattering.
The HERMES data have neither the coverage
nor the precision of E866, but provide a truly independent 
confirmation of the results.  Many of the systematic effects 
that are common to the NA51 and E866 Drell-Yan experiments do
not affect the HERMES measurement.  

These measurements of $\bar d(x)/\bar u(x)$ can be compared to 
the NMC DIS results by integrating $\bar d(x) - \bar u(x)$.  Table \ref{tab:idmupar}
summarizes the value of this integral over various 
$x$ ranges as parameterized by three global fits and as measured by E866.  
Table~\ref{tab:idmuexp}
summarizes three experimental determinations of this integral
over all $x$ values. The E866 integral is smaller than those
from NMC and HERMES, but consistent with them within the quoted uncertainties.

\begin{table}[tbp]
  \caption{$\int \left[\bar d(x) - \bar u(x)\right] dx $ evaluated over different
  $x$ ranges based on three parameterizations and as
  measured by E866 ($Q^2=54~\rm{GeV^2/c^2}$). }
  \label{tab:idmupar}
  \begin{center}
    \begin{tabular}{ccccc}
$x$ range      & CTEQ5M   & MRST   & GRV98  & E866 \\ \hline
$0<x<1$        & 0.1255   & 0.1149 & 0.1376 & 0.118$\pm$0.012   \\
$0.35<x<1$     &-0.0001   &-0.0003 & 0.0004 &    \\
$0.015<x<0.35$ & 0.0837   & 0.0815 & 0.0897 & 0.0803$\pm$0.011    \\
$0<x<0.015$    & 0.0418   & 0.0337 & 0.0475 &    \\
    \end{tabular}
  \end{center}
\end{table}

\begin{table}[tbp]
  \caption{$\int \left[\bar d(x) - \bar u(x)\right] dx $ as determined by three
  experiments.  The range of the measurement is shown along with the
  value of the integral over all $x$.}
  \label{tab:idmuexp}
  \begin{center}
    \begin{tabular}{ccc}
Experiment   & $x$ range       & $\int_0^1 \left[ \bar d(x) - \bar u(x)\right] dx $\\ \hline
E866         & $0.015<x<0.35$  & $0.118 \pm 0.012$ \\
NMC          & $0.004<x<0.80$  & $0.148 \pm 0.039$ \\
HERMES       & $0.020<x<0.30$  & $0.16  \pm 0.03$  \\
    \end{tabular}
  \end{center}
\end{table}

\section{Structure of the Nucleon Sea}

Ever since evidence for a flavor-asymmetric sea was reported by NMC
and NA51, the groups performing global analysis for 
PDFs have required $\bar d$ to be different
from $\bar u$. The NMC result constrains
the integral of $\bar d - \bar u$ to be $0.148 \pm 0.039$, while the NA51
result requires $\bar d / \bar u$ to be $ 1.96 \pm 0.25$ at $x = 0.18$.  Clearly,
the $x$-dependences of $\bar d - \bar u$ and $\bar d/ \bar u$ were
undetermined.
Recently, several PDF groups have published~\cite{CTEQ5,mrst,grv98} new parameterizations taking
into account new experimental results, including the E866 data reported in Ref.~\cite{prl}.
The parameterizations of the $x$ dependences of $\bar d - \bar u$ are now
strongly constrained by E866. 
As shown in Figure~\ref{fig:duwpdf}, these new 
parameterizations give significantly different shapes for $\bar d/ \bar u$
at $x > 0.15$ compared to previous works such as CTEQ4M and MRS(r2).

\begin{figure}
  \begin{center}
    \mbox{\epsfxsize=8.5cm\epsffile{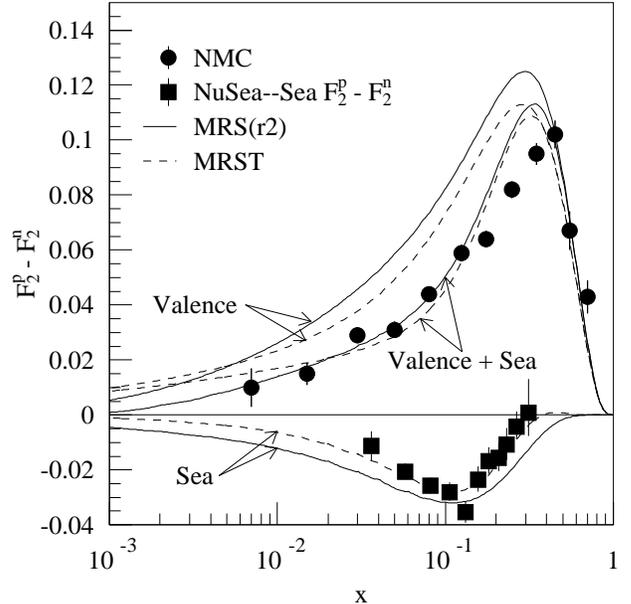}}  
  \end{center}                                                                     
\caption{$F^p_2 - F^n_2$ as measured by NMC at $Q^2$ = 4 \rm{GeV$^2/$c$^2$} compared with
next-to-leading-order calculations based on the MRS(r2) and MRST parameterizations.  Also
shown are the original E866 results from Ref.\protect \cite{prl}, scaled to $Q^2$ = 4 \rm{GeV$^2/$c$^2$}, for
the sea-quark contribution to $F^p_2
- F^n_2$. For each parameterization, the top (bottom) curve is the valence
(sea) contribution and the middle curve is the sum of the two.}
\label{fig:3.6.3}
\end{figure}

It is interesting to note that the E866 data
also affect the parameterization of the valence-quark distributions.
Figure~\ref{fig:3.6.3} shows the NMC data for $F_2^p
- F_2^n$ at $Q^2$ = 4~\rm{GeV$^2/$c$^2$}, together with the fits of MRS(r2) and
MRST. It is instructive to
decompose $F_2^p(x) - F_2^n(x)$ into contributions from 
valence and sea quarks:
\begin{eqnarray}
\lefteqn{F_2^p(x) -F_2^n(x) = }\hspace{0.75in} \nonumber \\
 & {1 \over 3} x \left[u_v(x) - d_v(x)\right] + {2 \over
3} x \left[\bar u(x) - \bar d(x)\right].
\end{eqnarray}
As shown in Fig.~\ref{fig:3.6.3}, the E866 data
provide a direct determination of the sea-quark contribution to $F_2^p
- F_2^n$. 
(The original E866 results from Ref.~\cite{prl} are shown, rather than the more precise 
results reported here, because they were used as inputs for the MRST PDF fits.)  
In order to preserve the fit to $F_2^p - F_2^n$, the MRST
parameterization for the valence-quark distributions, $u_v - d_v$,
is significantly lower in the region $x > 0.01$ than MRS(r2). Indeed, one of the major new
features of MRST is that $d_v$ is now significantly larger than before for
$x > 0.01$. Although the authors of MRST attribute this to the new 
$W$-asymmetry data from CDF and the new NMC results on $F_2^n/F_2^p$, it 
appears that the new information on $\bar d(x) - \bar u(x)$ also has a direct
impact on the valence-quark distributions.

\begin{figure}
  \begin{center}
    \mbox{\epsfxsize=8.5cm\epsffile{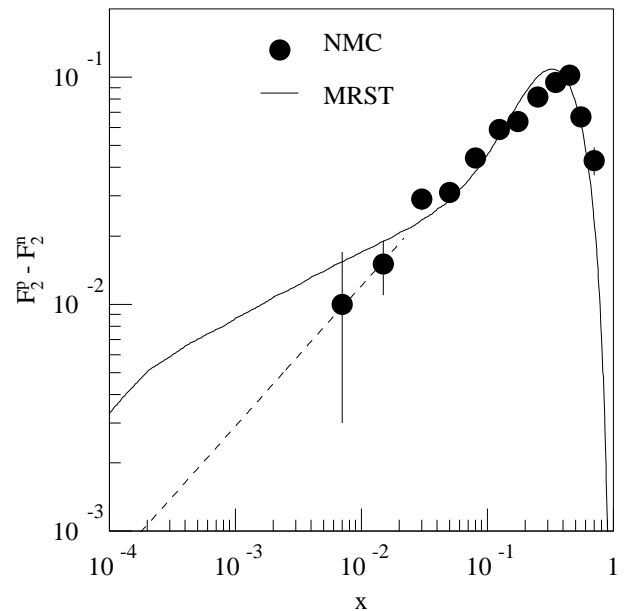}}                                     
  \end{center}                                                                     
\caption{$F^p_2 - F^n_2$ as measured by NMC at $Q^2$ = 4 \rm{GeV$^2/$c$^2$} compared with
the parameterization of MRST. The dashed curve corresponds to $0.21 x^{0.62}$,
a parameterization assumed by NMC for the unmeasured small-$x$ region
when the Gottfried integral was evaluated.}
\label{fig:3.6.4}
\end{figure}

Another implication of the E866 data is on the behavior of
$F_2^p - F_2^n$ at small $x$. In order to satisfy the constraint
$\int_0^1 [u_v(x) - d_v(x)] dx = 1$, the MRST values of $u_v(x) - d_v(x)$
at $x < 0.01$ are now much larger than in previous PDFs.
This is because the MRST parameterization of $u_v(x) -d_v(x)$
at $x > 0.01$ is smaller than before. As a consequence, $F_2^p - F_2^n$
is increased at small $x$ and MRST predicts a large contribution to the
Gottfried Sum from the small-$x$ ($x < 0.004$) region, as shown in
Fig.~\ref{fig:3.6.4}. If the MRST
parameterization for $F_2^p - F_2^n$ at $x < 0.004$ were used together
with the NMC data at $x > 0.004$, one would deduce a larger value for
the Gottfried Sum, and a value for the $\bar d - \bar u$
integral smaller than that of Eq. (3). This would bring better agreement 
between the E866 and the NMC
results on the $\bar d - \bar u$ integral.

\section{Origins of the Nucleon Sea}

\begin{figure}
  \begin{center}
    \mbox{\epsfxsize=8.5cm\epsffile{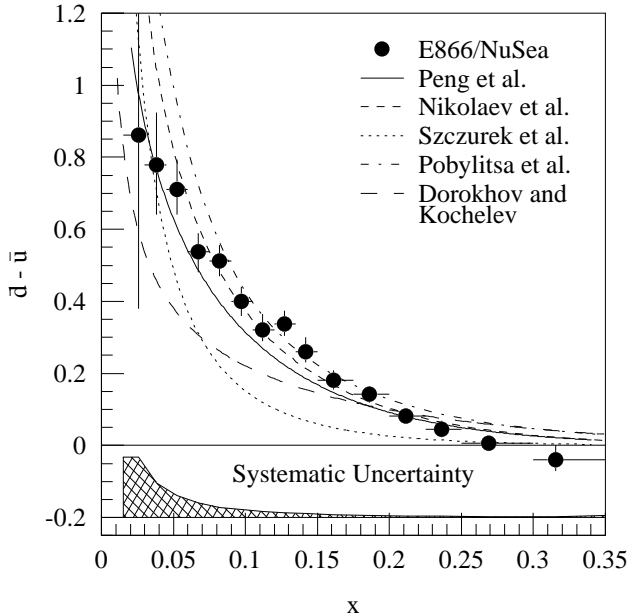}}
  \end{center}                                                                     
\caption{Comparison of the measured ${\bar d}(x)-{\bar u}(x)$ at $Q^2$ = 54
\rm{GeV$^2/$c$^2$} to predictions of
several models of the nucleon sea.  The solid and short-dash curves show pion-cloud
calculations by Peng {\it et al}.\@  and Nikolaev {\it et
al}.\@, respectively.  The dotted curve shows the chiral
perturbation theory calculation of Szczurek {\it et al}.\@,
while the dot-dash curve shows the chiral quark-soliton calculation of Pobylitsa 
{\it et al}.\@.   The long-dash curve shows the instanton
model prediction of Dorokhov and Kochelev.}
\label{fig:comp_mod}
\end{figure}

The FNAL E866/NuSea results for ${\bar d}(x)/{\bar u}(x)$ and ${\bar d}(x)-
{\bar u}(x)$ provide important constraints on models that attempt to describe the
origins of the nucleon sea and its antiquark asymmetry.  The early assumption of
flavor symmetry in the nucleon sea presumed that the primary mechanism to generate 
the sea is gluon splitting into $u{\bar u}$ and $d{\bar d}$ pairs.  
 Field and Feynman \cite{Fiel77} suggested that the extra
valence $u$ quark in the proton could lead to a suppression of $g \rightarrow
u{\bar u}$ relative to $g \rightarrow d{\bar d}$ via Pauli
blocking.  Ross and Sachrajda \cite{Ross79} subsequently calculated that the
effects of Pauli blocking are very small, and more recent calculations
\cite{Stef97} have confirmed this result, even indicating that the overall effect of
Pauli blocking may have the opposite effect to naive expectations. Given the small
mass difference between the $u$ and $d$ quarks, we are left with the conclusion
that perturbative QCD is incapable of generating a ${\bar d}/{\bar u}$ asymmetry
of the magnitude observed by E866.  Thus, this effect must have a 
non-perturbative origin.  
As these nonperturbative
mechanisms are considered, it is important to remember that
they act in addition to the perturbative sources, which
tends to dilute their effect. In effect the non-perturbative
sources must be even stronger to account for the large asymmetries
shown here.
Several models have been proposed, including 
meson-cloud contributions, chiral-field or chiral-soliton effects, and instantons.
Figure \ref{fig:comp_mod} compares the E866 results for 
$\bar{d}(x) - \bar{u}(x)$ to predictions of representative models of each of these types.

The coupling of the nucleon to virtual states containing isovector mesons provides
a natural mechanism to produce a ${\bar d}/{\bar u}$ asymmetry.  For example,
the valence quarks present in the $\pi N$ component of the proton have ${\bar
d}/{\bar u} = 5$.  Since Thomas pointed out their importance \cite{Thom83},
many authors have investigated virtual 
meson-baryon Fock states of the nucleon as the origin of the ${\bar d}/{\bar u}$
asymmetry in the sea.  Two recent reviews \cite{Kuma98,Spet98} provide a detailed 
survey of the literature.

Most calculations include contributions from $\pi N$ and $\pi \Delta$
configurations. $g_{\pi N N}$ and $g_{\pi N \Delta}$ are the well 
known pion-nucleon and pion-delta coupling constants, so the
primary difference among the various calculations is the treatment of 
the $\pi N N$
and $\pi N \Delta$ vertex form factors.  As an example, Fig.~\ref{fig:comp_mod}
compares the present determination of ${\bar d}(x)-{\bar u}(x)$ 
to  a pion-cloud-model calculation \cite{peng98}, which followed a procedure detailed by
Kumano \cite{Kuma91}.  In this calculation, dipole form factors were used, with
$\Lambda$
= 1.0~\rm{GeV} for the $\pi N N$ vertex and 
$\Lambda$ = 0.8~\rm{GeV} for the $\pi N \Delta$ vertex.  This calculation is
typical of many of this type, in that the probability of finding the nucleon in a
$\pi N$ configuration is approximately twice that of finding it in the $\pi \Delta$
configuration \cite{Szcz94,Koep96}.  However, a recent calculation by
Nikolaev {\it et al}.\@ \cite{Niko99}, also shown in 
Fig.~\ref{fig:comp_mod}, calls this into question.  After isolating the
contribution to inclusive particle production from Reggeon exchange, they
conclude that the $\pi N \Delta$ vertex should be substantially softer than
previously believed, significantly reducing the probability of finding the nucleon in
a $\pi \Delta$ configuration. 
  It adopts Gaussian form factors with cutoff parameters of 1 GeV$^{-2}$
  for the $\pi NN$ vertex and 2 GeV$^{-2}$ for the $\pi N \Delta$ vertex. This calculation predicts 
  that the $\pi N$ component of the nucleon is
  slightly more probable than in Ref.~\cite{peng98}
  and the $\pi \Delta$ component is very small. Thus,
  while it provides very good agreement with the E866 results for $x>0.05$, it
  contains significantly more singular behavior as $x \rightarrow 0$.
  Overall, it predicts that
  \begin{equation}
  \int_0^1 \, [\bar{d}(x)-\bar{u}(x)] \, dx = 0.177.
  \end{equation}

While the pion-cloud calculations above give a good description of the measured
${\bar d}(x)-{\bar u}(x)$, they 
are not able to predict ${\bar d}(x)/{\bar u}(x)$ since neither one attempts to describe the
entire light antiquark sea.  Rather, they assume that an additional symmetric contribution 
exists due to gluon splitting to bring
the ${\bar d}/{\bar u}$ ratio down to the measured value.
These models do however indicate that pions make up a large part of the sea where the asymmetry is greatest.  
In contrast, Alberg {\it
et al}.\@ \cite{Albe99} have investigated whether or not the entire light antiquark sea
might be understood in a meson-cloud picture.  They find that, by considering 
$\pi N$ and $\omega N$ contributions, they can fit ${\bar d}(x)-{\bar u}(x)$ and
simultaneously obtain a reasonable description of ${\bar d}/{\bar u}$ 
at $x<0.25$.  They also
speculate that the addition of $\pi \Delta$, $\rho N$ and $\sigma N$ terms would
preserve the fit to ${\bar d}-{\bar u}$, because of a cancellation between the $\pi
\Delta$ and $\rho N$ effects, and further improve the agreement for ${\bar
d}/{\bar u}$.

A different approach to the ${\bar d}/{\bar u}$ asymmetry, based on chiral
perturbation theory, has been proposed by Eichten {\it et al}.\@ \cite{Eich92}. 
Within their model, the asymmetry arises from the coupling of constituent quarks
to Goldstone bosons, such as $u \rightarrow d \pi^+$ and 
$d \rightarrow u \pi^-$. 
The excess of ${\bar d}$ over ${\bar u}$ is then simply due to the additional
valence $u$ quark in the proton.  Figure \ref{fig:comp_mod} includes the result of
such a calculation, based on a calculation of ${\bar d}(x)-{\bar u}(x)$ at $Q_0$ =
0.5~\rm{GeV/c} by Szczurek {\it et al}.\@ \cite{Szcz96}, and evolved to $Q^2 = 54~\rm{GeV^2/c^2}$. 
It clearly predicts too soft an asymmetry.  This arises because the model treats the
three valence quarks equivalently at the initial scale, with each carrying 
1/4  of the nucleon momentum.  (Gluons carry the remaining 1/4.)  
The ${\bar d}/{\bar u}$ ratio is then fixed by 
Clebsch-Gordan coefficients to be 11/7 for all $x$ at $Q_0$.  With this input,
QCD evolution requires ${\bar d}/{\bar u} \leq$ 11/7, independent of $x$ and
$Q$.  Hence, unlike
the meson-baryon models, this model underpredicts ${\bar d}/{\bar u}$ over much
of the measured $x$ range.  E866 results suggest that additional correlations
between the chiral constituents of the nucleon need to be taken into account.
The chiral quark-soliton model has been used by Pobylitsa {\it et al}.\@
\cite{Poby99} to calculate ${\bar d}(x)-{\bar u}(x)$ in the large-$N_c$ limit. 
Figure \ref{fig:comp_mod} shows that this model reproduces the measured ${\bar
d}(x)-{\bar u}(x)$ values well for $x>0.08$, but it overestimates the asymmetry at
small $x$. 

The spin and flavor structure of the nucleon sea have been
investigated in the instanton model by Dorokhov and Kochelev
\cite{instanton}.  They derive expressions for the $x$
dependence of the instanton-induced sea that are appropriate
for very large and very small $x$.  They then combine the
two asymptotic forms to obtain an {\it ad hoc} expression
for all $x$,\begin{equation}
\bar{d}_I(x) - \bar{u}_I(x) = 1.5\,A\,
\frac{(1-x)^7}{x\,\ln^2 x},
\end{equation}
where $A$ is an arbitrary constant which they chose to
reproduce early NMC results.  This form gives a poor 
description of our measured $\bar{d}(x)-\bar{u}(x)$,
as shown in Fig.~\ref{fig:comp_mod}, where we have set $A=0.163$ to give
$\int_0^1 (\bar{d}-\bar{u}) dx = 0.118$.  The model also predicts that
instanton-induced antiquarks should arise primarily at large
$p_T$ ($\langle {p_T}^2 \rangle \approx $ 2 \rm{GeV$^2 / c^2$}), but 
Fig.~\ref{fig:ratiopt} shows that the asymmetry we have
measured is not primarily a high-$p_T$ effect.  
Finally, the model
predicts that ${\bar d}/{\bar u} \rightarrow 4$ as $x \rightarrow 1$ for
the instanton-induced component of the nucleon sea.  Clearly, the 
experimental results strongly contradict this, so this model would
require a large additional contribution to the sea from $g \rightarrow
q{\bar q}$ as $x \rightarrow 1$ to bring ${\bar d}/{\bar u}$ into
agreement.   We do not
know if an alternative formulation of the instanton model,
especially including a more realistic treatment of the momentum dependence at 
finite $x$, might provide a better
description of our results.

\section{Conclusions}

While previous experiments have indicated that $\bar d >\bar u$, 
FNAL E866/NuSea was the first measurement of the $x$-dependence of
the flavor asymmetry in the nucleon sea.  This measurement has had 
an impact in several areas.  The global parameterizations 
of the nucleon sea have changed to fit these new data.  
Surprisingly, this measurement, when used in conjunction with the NMC
measurement, puts new and tighter constraints on the valence PDFs.
This measurement has also provided a means of testing the 
predictions of several nonperturbative models \cite{peng98}.  The 
unexpected sharp downturn in $\bar d(x)/\bar u(x)$ apparently back to unity
at the large $x$ limits of this measurement has prompted 
interest~\cite{thomas} in extending 
the measurement of $\sigma^{pd}/2\sigma^{pp}$ to higher $x$.  
An experiment has been proposed~\cite{p906} to make this measurement 
using the 120 \rm{GeV/c} proton beam from the new Main Injector at Fermilab.

The primary goal of this experiment was the determination
of $\sigma^{pd}/2\sigma^{pp}$ over a wide kinematic range.
The combined result from all three mass settings is 
shown in Fig.~\ref{fig:ratiowpdf} along with the curves from the 
calculated cross section ratio using various parameterizations.  
Parameterizations that do not 
include the first published results~\cite{prl} from this 
experiment do not agree well with the data.
>From the complete set of data, $\bar d(x)/\bar u(x)$,  
$\bar d(x) - \bar u(x)$, and $\int [\bar d(x) - \bar u(x)] dx$
of the proton were determined. These are shown in 
Figs.~\ref{fig:duwpdf}, \ref{fig:dmu}, and \ref{fig:idmu}.
Models that explicitly include pions in the proton
wavefunction~\cite{peng98} are relatively successful at reproducing the 
observed flavor asymmetry.

\acknowledgements

We would like to thank the Fermilab Particle Physics, Beams, and
Computing Divisions for their assistance in performing this
experiment.  
We would also like to thank W. K. Tung of the CTEQ collaboration for providing us 
with the code necessary to calculate the next-to-leading-order cross-section ratio.
This work was supported in part by the U.S. Department of Energy.

\end{document}